\documentclass[journal]{IEEEtran}

\usepackage{cite}
\usepackage{amsmath,amssymb,amsfonts}
\usepackage{algorithmic}
\usepackage{graphicx}
\usepackage{textcomp}
\usepackage{balance}
\usepackage{xcolor}
\def\BibTeX{{\rm B\kern-.05em{\sc i\kern-.025em b}\kern-.08em
		T\kern-.1667em\lower.7ex\hbox{E}\kern-.125emX}}
\usepackage[ruled,vlined,linesnumbered]{algorithm2e}
\usepackage{mathtools}
\usepackage[T1]{fontenc}
\usepackage{url}
\usepackage{dsfont}
\usepackage{color}
\usepackage{bm}

\usepackage{multirow}

\usepackage{pgfplotstable}
\usepackage{pgfplots}
\pgfplotsset{
	compat=newest,
	every tick label/.append style={font=\tiny},
	every axis plot/.append style={line width=0.5pt, mark size=1.5pt},	
	width=0.75\columnwidth,
	height=1.5cm,
}
\usetikzlibrary{plotmarks}
\usetikzlibrary{arrows.meta}
\usepgfplotslibrary{patchplots}
\usepackage{grffile}

\usepackage{tudacolors}
\usepackage{caption}	
\usepackage{subcaption}
\usepackage{diagbox}
\usepackage{framed}
\usepackage{CJKutf8} 
\usepackage{enumerate}

\usepackage{pgfplots}

\definecolor{MATLABcolor1}{rgb}{0.00000,0.44700,0.74100}%
\definecolor{MATLABcolor2}{rgb}{0.85000,0.32500,0.09800}%
\definecolor{MATLABcolor3}{rgb}{0.92900,0.69400,0.12500}%
\definecolor{MATLABcolor4}{rgb}{0.49400,0.18400,0.55600}%
\definecolor{MATLABcolor5}{rgb}{0.46600,0.67400,0.18800}%
\definecolor{MATLABcolor6}{rgb}{0.30100,0.74500,0.93300}%
\definecolor{MATLABcolor7}{rgb}{0.63500,0.07800,0.18400}%

\pgfplotscreateplotcyclelist{MATLABcolorlist}{
	MATLABcolor1, MATLABcolor2, MATLABcolor3, MATLABcolor4, MATLABcolor5, MATLABcolor6, MATLABcolor7
}

\pgfplotscreateplotcyclelist{MATLABcolormarklist}{
	{MATLABcolor1, mark=o, every mark/.append style = {solid}}, 
	{MATLABcolor2, mark=diamond, every mark/.append style = {solid}}, 
	{MATLABcolor3, mark=triangle, every mark/.append style = {solid}}, 
	{MATLABcolor4, mark=star, every mark/.append style = {solid}}, 
	MATLABcolor5, MATLABcolor6, MATLABcolor7
}

\definecolor{light-gray}{gray}{0.65}

\newcommand{\vect}[1]{\ensuremath{\mathbf{#1}}}

\newcommand\bb{\ensuremath{\mathbf{b}}}
\newcommand\bc{\ensuremath{\mathbf{c}}}
\newcommand\bd{\ensuremath{\mathbf{d}}}
\newcommand\be{\ensuremath{\mathbf{e}}}

\newcommand\br{\ensuremath{\mathbf{r}}}
\newcommand\bs{\ensuremath{\mathbf{s}}}

\newcommand\bv{\ensuremath{\mathbf{v}}}
\newcommand\bw{\ensuremath{\mathbf{w}}}
\newcommand\bx{\ensuremath{\mathbf{x}}}
\newcommand\by{\ensuremath{\mathbf{y}}}
\newcommand\bz{\ensuremath{\mathbf{z}}}

\newcommand\bA{\ensuremath{\mathbf{A}}}
\newcommand\bB{\ensuremath{\mathbf{B}}}

\newcommand\bD{\ensuremath{\mathbf{D}}}

\newcommand\bF{\ensuremath{\mathbf{F}}}

\newcommand\bH{\ensuremath{\mathbf{H}}}
\newcommand\bI{\ensuremath{\mathbf{I}}}

\newcommand\bP{\ensuremath{\mathbf{P}}}

\newcommand\bU{\ensuremath{\mathbf{U}}}
\newcommand\bV{\ensuremath{\mathbf{V}}}

\newcommand\bX{\ensuremath{\mathbf{X}}}
\newcommand\bY{\ensuremath{\mathbf{Y}}}
\newcommand\bZ{\ensuremath{\mathbf{Z}}}

\newcommand\bzero{\ensuremath{\mathbf{0}}}

\newcommand\bSigma{\ensuremath{\boldsymbol{\Sigma}}}

\newcommand{\Rbb}{\ensuremath{\mathbb{R}}}
\newcommand{\Cbb}{\ensuremath{\mathbb{C}}}

\newcommand{\Kbb}{\ensuremath{\mathbb{K}}}

\newcommand{\set}[1]{\ensuremath{\mathcal{#1}}}
\newcommand{\opr}[1]{\ensuremath{\mathcal{#1}}}

\DeclareMathOperator{\rank}{rank}
\DeclareMathOperator{\sign}{sign}

\DeclareMathOperator{\tr}{tr}

\newcommand\imagunit{\ensuremath{\mathrm{j}}}
\newcommand\euler{\ensuremath{\mathrm{e}}}

\newcommand{\norm}[1]{\lVert {#1} \rVert}
\newcommand{\Norm}[1]{\left\lVert {#1} \right\rVert}
\newcommand{\abs}[1]{\lvert {#1} \rvert}
\newcommand{\Abs}[1]{\left\lvert {#1} \right\rvert}

\newcommand{\tT}{\ensuremath{{\mathsf{T}}}}
\newcommand{\tH}{\ensuremath{{\mathsf{H}}}}
\newcommand{\tF}{\ensuremath{\mathsf{F}}}
\newcommand{\st}{\text{s.t.}}

\renewcommand{\vec}{\ensuremath{\operatorname{vec}}}

\newtheorem{thm}{Theorem}

\newtheorem{defi}[thm]{Definition}

\newcommand{\R}{\mathbb{R}}

\newcommand{\C}{\mathbb{C}}

\newcommand{\K}{\mathbb{K}}

\newcommand{\define}{=}

\newcommand{\x}{\bm{x}}

\algsetup{indent=1.5em}

\begin{document}

\title{Extended Successive Convex Approximation for Phase Retrieval with Dictionary Learning}

\author{Tianyi Liu \begin{CJK*}{UTF8}{gbsn}
		(刘添翼),
	\end{CJK*} Andreas M. Tillmann, Yang Yang \begin{CJK*}{UTF8}{gbsn}
	(杨阳),
\end{CJK*} Yonina C. Eldar, and Marius Pesavento
\thanks{Tianyi Liu and Marius Pesavento are with TU Darmstadt, 64283 Darmstadt, Germany (e-mail: tliu@nt.tu-darmstadt.de; pesavento@nt.tu-darmstadt.de).
This work was supported by the EXPRESS project within the DFG priority program CoSIP (DFG-SPP 1798) 
and the project ``Open6GHub'' (grant no. 16KISK014) sponsored by the Federal Ministry of Education and Research of Germany.
}
\thanks{Andreas M. Tillmann is with TU Braunschweig, 38106 Braunschweig, Germany (e-mail: a.tillmann@tu-braunschweig.de).}
\thanks{Yang Yang is with	Meta (\mbox{e-mail:} yangyang22@meta.com).}
\thanks{Yonina C. Eldar is with Weizmann Institute of Science, Rehovot 7610001, Israel (e-mail: yonina.eldar@weizmann.ac.il).}
\thanks{Part of this work is accepted for publication at IEEE ICASSP 2021~\cite{liuParallelAlgorithmPhase2021}.}
\thanks{Digital Object Identifier 10.1109/TSP.2022.3233253}
\thanks{\textcopyright 2023 IEEE.  Personal use of this material is permitted.  Permission from IEEE must be obtained for all other uses, in any current or future media, including reprinting/republishing this material for advertising or promotional purposes, creating new collective works, for resale or redistribution to servers or lists, or reuse of any copyrighted component of this work in other works.}
}

\maketitle

\begin{abstract}
	Phase retrieval aims at recovering unknown signals from magnitude measurements of linear mixtures. 
	In this paper, we consider the phase retrieval with dictionary learning problem, which includes another prior information that the signal admits a sparse representation over an unknown dictionary. The task is to jointly estimate the dictionary and the sparse representation from magnitude-only measurements.
	To this end, we study two complementary formulations and develop efficient parallel algorithms by extending the successive convex approximation framework using a smooth majorization. The first algorithm is termed \textit{compact-SCAphase} and is preferable in the case of moderately diverse mixture models with a low number of mixing components. It adopts a compact formulation that avoids auxiliary variables. The proposed algorithm is highly scalable and has reduced parameter tuning cost. The second algorithm, referred to as \textit{SCAphase}, uses auxiliary variables and is favorable in the case of highly diverse mixture models. It also renders simple incorporation of additional side constraints.
	The performance of both methods is evaluated when applied to blind channel estimation from subband magnitude measurements in a multi-antenna random access network.
	Simulation results show the efficiency of the proposed techniques compared to state-of-the-art methods.
	
\end{abstract}


\section{Introduction}
\label{sec:intro}

\IEEEPARstart{P}{hase} retrieval refers to the problem of recovering unknown signals from the (squared) magnitude of linear measurements corrupted by additive noise.
It has received considerable attention in various applications such as diffraction imaging~\cite{candesPhaseRetrievalCoded2015,shechtmanPhaseRetrievalApplication2015}, astronomy~\cite{fienupPhaseRetrievalAlgorithms1982}, and X-ray crystallography~\cite{harrisonPhaseProblemCrystallography1993}, where the measurement of intensity is much easier than that of phase. In some other applications, including non-coherent direction-of-arrival estimation~\cite{kimNoncoherentDirectionArrival2014}, the loss of phase information is caused by imperfect phase synchronization.

In recent years, numerous phase retrieval approaches have been developed, which can be principally classified as nonconvex and convex ones. 
In the nonconvex optimization methods, the recovery problem is formulated as a nonconvex least-squares (LS) problem. Stationary points of the nonconvex formulation can then be obtained by classic continuous optimization algorithms such as alternating projections~\cite{gerchbergPracticalAlgorithmDetermination1972,fienupPhaseRetrievalAlgorithms1982}, gradient descent~\cite{candesPhaseRetrievalWirtinger2015,wangSolvingSystemsRandom2018,chenGradientDescentRandom2019}, and alternating
direction method of multipliers (ADMM)~\cite{wenAlternatingDirectionMethods2012,liangPhaseRetrievalAlternating2017}.
A popular class of convex optimization approaches employs semidefinite relaxation~\cite{candesPhaseliftExactStable2013,candesPhaseRetrievalMatrix2015,waldspurgerPhaseRecoveryMaxCut2015,jaganathanSTFTPhaseRetrieval2016}, which lifts the problem to a higher dimension and is, hence, computationally prohibitive for large-scale problems. Recently, some non-lifting convex optimization approaches have been developed based on solving a basis pursuit problem in the dual domain, including PhaseMax~\cite{goldsteinPhaseMaxConvexPhase2018} and PhaseEqual~\cite{wangPhaseEqualConvexPhase2020}.
A comprehensive review of recent advances in phase retrieval from a numerical perspective is presented in~\cite{fannjiangNumericsPhaseRetrieval2020}.

On the other hand, additional prior information on the unknown signal, such as sparsity, can be used to improve uniqueness and stability of the reconstruction~\cite{eldarRecentAdvancesPhase2016}.
Most of the aforementioned phase retrieval approaches have been adapted to recovering signals that are sparse either in the standard basis or in a known dictionary~\cite{shechtmanGESPAREfficientPhase2014,eldarSparsePhaseRetrieval2014,qiuUndersampledSparsePhase2017,wangSparsePhaseRetrieval2018,pauwelsFienupMethodsSparse2018,salehiLearningPhaseRegularized2018,yangParallelCoordinateDescent2019,wangPhaseEqualConvexPhase2020}.
The GESPAR algorithm is based on the damped Gauss-Newton method~\cite{shechtmanGESPAREfficientPhase2014}. Majorization-Minimization (MM) algorithms are devised in~\cite{qiuUndersampledSparsePhase2017}. In~\cite{wangSparsePhaseRetrieval2018}, the Truncated Amplitude Flow (TAF) method is extended to recovering sparse signals. The STELA algorithm proposed in~\cite{yangParallelCoordinateDescent2019} is based on the successive convex approximation (SCA) and can be parallelized.

Phase retrieval was generalized in~\cite{tillmannDOLPHInDictionaryLearning2016} to jointly learning an unknown dictionary and a sparse representation. To tackle the joint estimation problem, the authors propose a regularized nonconvex LS formulation with squared magnitude measurements and develop an alternating minimization algorithm termed DOLPHIn. In~\cite{qiuUndersampledSparsePhase2017}, the authors apply a similar regularized LS formulation to magnitude measurements and solve it by an algorithm based on block successive upper-bound minimization (BSUM), named SC-PRIME. There, it is shown by both theoretical justification and numerical results that the reconstruction from magnitude measurement outperforms that from intensity measurements.
However, the use of auxiliary variables in both aforementioned methods depresses the scalability and, more notably, increases the number of hyperparameters that require tuning. Moreover, neither of the two methods can take full benefit of modern parallel hardware architectures. In addition, SC-PRIME often suffers from slow convergence due to the loose approximation used by BSUM.

The SCA framework~\cite{yangUnifiedSuccessivePseudoconvex2017} possesses the advantage of parallelism. However, it can only be applied to a composite problem with a smooth loss function and a convex but not necessarily smooth regularization. 
In this paper, to address the phase retrieval with dictionary learning problem given the magnitude measurements, which is formulated as a nonsmooth and nonconvex LS problem, we extend the SCA framework using a smooth majorization. The extended framework inherits the parallel nature of the original SCA framework.
Two efficient parallel algorithms for the phase retrieval and dictionary learning problem are proposed by applying the extended SCA framework to two complementary formulations, respectively.
Specifically, we first study a compact formulation that avoids the auxiliary variables and the proposed extended-SCA algorithm is termed \textit{compact-SCAphase}.
Then another algorithm based on the extended SCA framework is proposed for the conventional formulation with auxiliary variables, which is referred to as \textit{SCAphase} (Successive Convex Approximation for phase retrieval with dictionary learning).
The performance of the proposed algorithms is evaluated when applied to blind sparse channel estimation from subband magnitude measurements in a multi-antenna random access network.
Simulation results on synthetic data show the fast convergence of the proposed algorithms compared to the state-of-the-art method SC-PRIME~\cite{qiuUndersampledSparsePhase2017}. In the case with less diverse linear mixing models, compact-SCAphase is more competitive than SCAphase in terms of both computational complexity and parameter tuning cost. However, for highly diverse linear measurement operators, the computational complexity of compact-SCAphase dramatically grows, compared to SCAphase.
To summarize, the main contributions of this paper are:
\begin{itemize}
	\item We introduce an extension of the SCA framework for the phase retrieval with dictionary learning problem. Two efficient parallel algorithms are proposed by applying the extended SCA framework to two complementary formulations, respectively.
	\item The convergence of the extended SCA framework is established based on a generalized concept of stationarity. Our novel convergence analysis can also be used to establish the convergence of SC-PRIME~\cite{qiuUndersampledSparsePhase2017}.
	\item To reduce the overall computational complexity of compact-SCAphase, an efficient procedure based on rational approximation is devised for solving the $ \ell_2 $-norm constrained LS subproblems.
\end{itemize}

The paper is organized as follows. In Section~\ref{sec:problem}, we introduce the signal model and provide two different mathematical formulations with and without auxiliary variables, respectively, for the phase retrieval with dictionary learning problem. The proposed algorithms
for both formulations are described in Section~\ref{sec:alg1} and~\ref{sec:alg2}, respectively. 
In Section~\ref{sec:convergence&complexity}, we establish the convergence of the proposed algorithms and analyze the computational complexity in comparison to SC-PRIME.
Simulation results on synthetic data are presented and discussed in Section~\ref{sec:results} and conclusions are drawn in Section~\ref{sec:conclusion}.

\section{Notation and Problem Formulation}
\label{sec:problem}

\subsection{Notation}
We use $ x $, $ \bx $ and $ \bX $ to denote a scalar, column vector and matrix, respectively.
For any $x \! \in \! \Cbb$, $ \abs{x} $ denotes its magnitude, $\arg (x)$ its phase, $ \bar{x} $ its complex conjugate, and $ \Re (x) $ its real part.
The soft-thresholding operator is denoted by ${\cal S}_\lambda ( x ) \!=\! \max\{0,\Abs{x}-\lambda\}\cdot \euler ^{\imagunit \arg (x)}$.
Symbols $ (\cdot)^\tT$, $(\cdot)^\tH$, $(\cdot)^{-1} $, and $ (\cdot)^\dagger $ denote the transpose, Hermitian transpose, inverse and pseudoinverse, respectively.
For a matrix $\bX \!\in\! \Cbb^{M \! \times \! N}$, $x_{k,l}$ is its $(k,l)$th element, $\bx_l$ its $l$th column, $\bx_{k:}^\tT$ its $k$th row,
$ \vec (\bX) \!=\! [\bx_1^\tT,\ldots,\bx_N^\tT]^\tT $ its vectorized form, and $ \norm{\bX}_1 \!=\! \sum_{k,l}  \abs{x_{k,l}} $ its elementwise $\ell_1$-norm. 
The trace operator is written as $\tr(\cdot)$ and $ \norm{\cdot}_\tF $ is the Frobenius norm.
For a vector $\x$, the $k$th entry is $x_k$. Also, the $k$th entry of a vector $\x_l$ that itself carries a subscript will be denoted by $x_{k,l}$.
The Hadamard and Kronecker products are denoted by $\odot$ and $\otimes$, respectively; 
symbol $ \bzero $ is a zero matrix.

For a linear operator $\opr{F}(\cdot)$, $\opr{F}^*(\cdot)$ is its adjoint operator.
For a real-valued function $ f(\bX) $ with real arguments $ \bX \! \in \! \Rbb^{M \times N}$, $ \nabla_{\! \bX} f(\bX) $ is the gradient with respect to $\bX$, i.e., an $M \!\times\! N$ matrix with the $ (k,l) $th entry being $ \frac{\partial}{\partial x_{k,l}} \! f(\bX) $, and $ \nabla_{\!\! \vec (\bX)} f(\bX) $ is the vectorized form of the gradient. The Hessian defined with the vectorized arguments is denoted by  $ \nabla_{\!\! \vec (\bX)}^2 f(\bX) $, which is an $ MN \! \times \! MN $ matrix with the $ (m+(n-1)M,k+(l-1)M) $th entry being $ \frac{\partial^2}{\partial x_{m,n} \partial x_{k,l}} \! f(\bX)$. Also,~$ \nabla_{\!\! x_{k,l}}^2 f(\bX)$ is the diagonal entry corresponding to $ x_{k,l} $ in the Hessian.
For complex arguments $\bX \in \Cbb^{M \times N}$, the entries of gradient and Hessian are defined as 
$ [\nabla_{\! \bX} f(\bX)]_{k,l} \!=\! 2 \frac{\partial}{\partial \bar{x}_{k,l}} \! f(\bX) $ and $ [\nabla_{\!\! \vec (\bX)}^2 f(\bX)]_{m + (n-1)M,k + (l-1)M} \!=\! 2 \frac{\partial^2}{\partial \bar{x}_{m,n} \partial x_{k,l}} \! f(\bX)$, where $\frac{\partial}{\partial x}$ and $\frac{\partial}{\partial \bar{x}}$ are the Wirtinger derivative operators~\cite{spiegelComplexVariables2009}.
Thus, for both real and complex arguments,
a real-valued quadratic function $f(\bX)$ can be written as the quadratic Taylor series at any $ \vect{X}_0 $ in a unified form
	$ f(\bX) = f(\bX_0) + \Re \left( \tr \big( \Delta \bX^\tH \nabla_{\! \bX} f(\bX_0) \big) \right)
	+ \tfrac{1}{2} {\vec (\Delta \bX)}^\tH \nabla_{\!\! \vec (\bX)}^2 f(\bX_0) \vec (\Delta \bX) $
with $ \Delta \bX = \bX - \bX_0 $.

\subsection{Problem Formulation}

We consider the following nonlinear system. For an input signal $\bX \in \K^{N \times I}$,
$\K\in\{\R,\C\}$, the following noise-corrupted magnitude-only measurements are observed:
\begin{equation}
	\label{eq:measurements}
	\bY = \abs{\opr{F} (\bX)}+\vect{N},
\end{equation}
where $\opr{F}: \Cbb^{N \times I} \rightarrow \Cbb^{M_1 \times M_2}$ is a linear operator, $ \vect{N} $ is a noise matrix, and the absolute value operation $ \abs{\cdot} $ is applied elementwise.
The negative entries of $\bY$ caused by noise will be set to $0$.
A general linear mixing operator $ \opr{F} (\bX) $ can be written as
\begin{equation}
	\label{eq:oprF_general}
	\opr{F} (\bX) \!=\! \begin{matrix}
		\sum_{k=1}^K \bA_k \bX \bB_k
	\end{matrix}, 
\end{equation}
where $\bA_k \in \Cbb^{M_1 \times N}$ and $ \bB_k \in \Cbb^{I \times M_2}$, $ k = 1,\ldots,K $, perform the row and column mixing, respectively, and the number of distinct mixing components $K$ is termed as the diversity of the mixing operator $ \opr{F} $ in this paper.
Note that the linear operator $\opr{F}$ in~\eqref{eq:oprF_general} can be written equivalently in a vectorized form
\begin{equation}
	\label{eq:linearOpr_vectorized}
	\vec \big(\opr{F} (\bX) \big) = \bF \cdot \vec (\bX) \quad \text{with} \ \bF = \begin{matrix}
		\sum_{k=1}^{K} \bB_k^\tT \otimes \bA_k
	\end{matrix}, 
\end{equation}
which we will also use in this paper. 
Moreover, each column $\bx_i$ of $\bX$ is assumed to admit a sparse representation over an unknown dictionary $\bD \! \in \! \Kbb^{N \times P}$, i.e., $ \bx_i \!=\! \bD \bz_i $ with a sparse code vector $ \bz_i \in \Kbb^P $.
Let $ \bZ \!=\! [\bz_1,\ldots,\bz_{I}] $ summarize the code vectors. 
Our objective is to jointly learn the dictionary $ \bD $ and the sparse codes $ \bZ $ so as to minimize the (LS) reconstruction error. 

To this end, we solve the following compact formulation for phase retrieval with dictionary learning (cPRDL) problem:
\begin{equation}
	\label{prob:PRDLMatrixForm}
	\text{cPRDL:} \quad \underset{\bD \in \set{D}, \bZ}{\min}  \quad \tfrac{1}{2} \norm{\bY - | \opr{F}(\bD \bZ)|}^2_\tF + \lambda \norm{\bZ}_1.
\end{equation}
The first term evaluates the data fidelity
by the LS criterion, which is nonsmooth and nonconvex due to the absolute value operation.
The second term promotes sparsity in $\bZ$ with a regularization parameter $\lambda \! \geq \! 0$.
To avoid scaling ambiguities, we restrict $\bD$ to be in the convex set $\set{D} \!=\!  \{\bD \!\in\! \Kbb^{N \! \times \! P} \mid \norm{\bd_p}_2 \! \leq \! 1 \ \forall p \! =\! 1, \ldots, P \} $. Each column $\bd_p$ is called an atom and the dictionary size must be below the number of columns in $ \bX $, i.e., $ P \!<\! I $. Otherwise, each column $\bx_i$ can be trivially represented by a $ 1 $-sparse vector $\bz_i$ with an atom $\bx_i / \norm{\bx_i}_2$.


An alternative formulation for phase retrieval with dictionary learning (PRDL), which we will also consider, is constructed as follows with an auxiliary variable~$\bX$:
\begin{equation}\label{prob:SC-PRIME}
	\text{PRDL:} \quad \underset{\bX, \bD \in \set{D}, \bZ}{\min} \ \tfrac{1}{2} \norm{\bY \!-\! | \opr{F}(\bX)|}^2_\tF \!+\! \tfrac{\mu}{2} \norm{\bX \!-\! \bD \bZ}^2_\tF \!+\! \rho \norm{\bZ}_1. 
\end{equation}
The additional second term measures how well the signal $\bX$ can be approximated by the sparse representation $\bD \bZ$. Two regularization parameters $ \mu \geq 0$ and $\rho \! \geq \! 0 $ are used to balance the data fidelity, the approximation quality, and the code sparsity.


The formulation~\eqref{prob:SC-PRIME} was first proposed in~\cite{tillmannDOLPHInDictionaryLearning2016},
however, with the intensity measurements 
$ \widetilde{\bY} \!=\! \abs{\opr{F}(\bX)}^2 \!+\! \vect{N} $,
which results in another smooth data fidelity term $ \tfrac{1}{4} \norm{\widetilde{\bY} - \abs{\opr{F}(\bX)}^2}_\tF^2 $. 
In~\cite{qiuUndersampledSparsePhase2017}, the authors have shown that, for the intensity measurements $ \widetilde{\bY} $, it is also beneficial, in the high SNR regime, to use formulation~\eqref{prob:SC-PRIME}
with the modulus information $ \sqrt{\widetilde{\bY}} $, where $\sqrt{\cdot}$ is applied elementwise, due to the reduced noise level in $ \sqrt{\widetilde{\bY}} $. 
Thus, we consider the magnitude measurement model~\eqref{eq:measurements}.

In~\cite{qiuUndersampledSparsePhase2017}, the state-of-the-art SC-PRIME algorithm is devised for the conventional formulation~\eqref{prob:SC-PRIME} based on BSUM, which, however, does not take full advantage of modern parallel hardware architectures. Also, the conservative majorization in SC-PRIME often results in slow convergence. Therefore, we develop the \textit{compact-SCAphase} and \textit{SCAphase} algorithms for the compact formulation~\eqref{prob:PRDLMatrixForm} and conventional formulation~\eqref{prob:SC-PRIME}, respectively, based on an extension of SCA framework.
Both proposed algorithms can be easily parallelized.

The two proposed algorithms are advantageous in different scenarios. The conventional formulation~\eqref{prob:SC-PRIME} is not suitable for large-scale problems due to the introduction of auxiliary variables. Also, the complexity of tuning two regularization parameters $\mu$ and $\rho$ in~\eqref{prob:SC-PRIME} is significantly higher than that of tuning one parameter. However, compared to SCAphase, the computational complexity of compact-SCAphase grows dramatically with the increase of diversity of the designed linear measurement operator $\opr{F}$. Moreover, the conventional formulation~\eqref{prob:SC-PRIME} admits simple incorporation of additional prior information on $ \bX $ such as nonnegativity in radio astronomy~\cite{fienupPhaseRetrievalAlgorithms1982}.

In the following, we describe the proposed compact-SCAphase and SCAphase algorithms. The derivations are based on the model with complex-valued variables. However, the same derivations can be made for the real-valued case.

\section{Proposed Algorithm for Formulation cPRDL}
\label{sec:alg1}

In this section, by extending the SCA framework in~\cite{yangUnifiedSuccessivePseudoconvex2017,yangInexactBlockCoordinate2020}, we propose an efficient iterative algorithm to find a stationary point of~\eqref{prob:PRDLMatrixForm} via a sequence of approximate problems that can be solved in parallel. 
We denote the objective function in~\eqref{prob:PRDLMatrixForm} by $h(\bD,\bZ) = f(\bD,\bZ) + g(\bZ)$
with
\begin{equation}
	\label{eq:f+g}
	f(\bD,\bZ) = \tfrac{1}{2} \norm{\bY - \abs{\opr{F}(\bD \bZ)}}^2_\tF \quad \text{and} \quad g(\bZ) = \lambda \norm{\bZ}_1.
\end{equation}
The problem is challenging since $g$ is nonsmooth and, more notably, $f$ is nonsmooth and nonconvex. 
To overcome this difficulty, in each iteration, we first majorize $f$ by a smooth function, which naturally leads to a majorization for the overall objective function $h$.
Then the majorizing function is only minimized approximately. In particular, we obtain a descent direction of the majorizing function by minimizing exactly its convex approximation. The variable can then be updated along this descent direction with a suitable step size, which can be efficiently obtained by exact line search. Consequently, a decrease of the original objective function $h$ is also ensured.

From the procedure described above, it can be noticed that the convergence of the proposed compact-SCAphase algorithm cannot be established under the framework of MM or SCA since the gradient consistency condition~\cite[A2.2]{sunMajorizationMinimizationAlgorithmsSignal2017} is apparently not satisfied.
Nonetheless, in Section~\ref{subsec:convergence}, we prove that compact-SCAphase converges to a stationary point of problem~\eqref{prob:PRDLMatrixForm} according to a generalized concept of stationarity.

Once a stationary point $ (\bD^\star, \bZ^\star) $ of the cPRDL problem in~\eqref{prob:PRDLMatrixForm} has been obtained by the compact-SCAphase algorithm, we optionally perform a debiasing step similar to that in~\cite{figueiredoGradientProjectionSparse2007} to further improve the estimation quality, which solves an instance of the cPRDL problem with $\lambda = 0$ and a restriction that the entries $ z_{p,i} $ having zero values in $ \bZ^\star $ are fixed at zero.


\subsection{Smooth Majorization}
\label{subsec:approx_1}

We first derive a smooth majorizing function for $f$ in~\eqref{eq:f+g} by following a similar approach as in~\cite{qiuUndersampledSparsePhase2017}.
Let $ \vect{S} = (\vect{D},\vect{Z}) $ denote the collection of all variables,
and let $\vect{S}^{(t)} = (\bD^{(t)},\bZ^{(t)})$ be the current point at iteration $t$. Also,  Function $f$ can be expanded as
$ f(\vect{S}) \!=\! \tfrac{1}{2} ( \norm{\bY}^2_\tF \!+\! \norm{ \opr{F} (\bD \bZ) }^2_\tF ) \!-\! \tr  (\bY^\tH \abs{\opr{F} (\bD \bZ)} ) $.
We note that
\begin{equation}
	\label{eq:upbound}
	\Abs{x} = \abs{x \cdot \euler^{\imagunit \phi}} \geq \Re  (x \cdot \euler^{\imagunit \phi}) \quad \text{for any } x \in \Cbb \text{ and } \phi \in [0, 2 \pi),
\end{equation}
and that equality holds for $\phi = - \arg(x)$.
Defining 
$ \bY^{(t)} = \bY \odot \euler^{\imagunit \arg ( \opr{F} (\bD^{(t)} \bZ^{(t)}) )}  $,
where $\euler^{(\cdot)}$ and $\arg (\cdot)$ are applied elementwise,
we construct the following function:
\begin{multline}
	\label{eq:PRDLproblemFctUpper}
	\widehat{f} (\vect{S}; \vect{S}^{(t)}) 
	 = - \tr \left( \bY^\tH \Re \big( \opr{F} (\bD \bZ) \odot \euler^{-\imagunit \arg (\opr{F} (\bD^{(t)} \bZ^{(t)}))} \big) \right) \\
	+ \tfrac{1}{2} ( \norm{\bY}^2_\tF + \norm{\opr{F} (\bD \bZ)}^2_\tF )
	 = \tfrac{1}{2} \norm{{\bY}^{(t)} - \opr{F} (\bD \bZ) }_\tF^2.
\end{multline}
As $ \bY $ contains nonnegative entries, we can infer from~\eqref{eq:upbound} that
	$ \widehat{f} (\vect{S}^{(t)};\vect{S}^{(t)}) = f(\vect{S}^{(t)}) $ and
	$ \widehat{f} (\vect{S};\vect{S}^{(t)}) \geq f(\vect{S})$ for all $ \vect{S} $.
Thus, $\widehat{f}(\vect{S};\vect{S}^{(t)})$ is a smooth majorizing function of $f$ at point $\vect{S}^{(t)}$~\cite{hunterTutorialMMAlgorithms2004,sunMajorizationMinimizationAlgorithmsSignal2017},
which has the partial gradients
\begin{equation}
	\label{eq:grad_upperBoundf}
	\begin{aligned}
		\nabla_{\! \bD} \widehat{f} (\vect{S};\vect{S}^{(t)}) &= \opr{F}^* \big( \opr{F}(\bD\bZ) - \bY^{(t)} \big) \cdot \bZ^\tH \\
		\text{and }\nabla_{\! \bZ} \widehat{f} (\vect{S};\vect{S}^{(t)}) &= \bD^\tH \cdot \opr{F}^* \big( \opr{F}(\bD\bZ) - \bY^{(t)} \big).
	\end{aligned}
\end{equation}
However, $\widehat{f}$ is nonconvex due to the bilinear map $\bD \bZ$.
Then function
$ \widehat{h} (\vect{S};\vect{S}^{(t)}) = \widehat{f} (\vect{S};\vect{S}^{(t)}) + g(\bZ) $
is a majorizing function of the objective function $h$ at $\vect{S}^{(t)}$.

\subsection{Separable Convex Approximation}

Next, departing from the classic MM algorithm~\cite{hunterTutorialMMAlgorithms2004,sunMajorizationMinimizationAlgorithmsSignal2017}, where $\widehat{h}$ is minimized exactly at a high computational cost, 
we further construct a convex approximate problem
that can be decomposed into subproblems and solved in parallel.


As the regularization $g$ is convex and separable, we leave $ g $ unaltered and only design a separable convex approximation for $\widehat{f}$ at the current point $ \vect{S}^{(t)} $. As $\widehat{f}$ is partially convex in $\bD$ and $\bZ$, respectively, we adopt the best-response approximation, where the approximate function is the sum of several components~\cite{yangUnifiedSuccessivePseudoconvex2017}. In each component, only part of the variables are varied while the rest are fixed to their current values.
Let $\widetilde{f}_D (\bD;\vect{S}^{(t)})$ and $\widetilde{f}_Z (\bZ;\vect{S}^{(t)})$ be the approximate functions of $\widehat{f}(\vect{S};\vect{S}^{(t)})$ over $\bD$ and~$\bZ$, respectively. 
They are devised as
\begin{equation}	\label{eq:PRDLapprox_f}
	\begin{aligned}
		\widetilde{f}_D (\bD;\vect{S}^{(t)}) &=  \begin{matrix}
			\sum_{p=1}^{P} \widehat{f} ( \bd_p,\bD^{(t)}_{-p},\bZ^{(t)}; \vect{S}^{(t)} )
		\end{matrix},\\
		\widetilde{f}_Z (\bZ;\vect{S}^{(t)}) &=  \begin{matrix}
			\sum_{i=1}^{I}\sum_{p=1}^{P} \widehat{f} (z_{p,i},\bD^{(t)},\bZ^{(t)}_{-(p,i)}; \vect{S}^{(t)} )
		\end{matrix},
	\end{aligned}
\end{equation}
where $ \bD_{-p} \in \Cbb^{N \times (P-1)} $ is obtained by removing $ \bd_p $ from $\bD$ and $ \bZ_{-(p,i)} $ the collection of all entries of $\bZ$ except $z_{p,i}$.
Then the convex approximation of $\widehat{h} (\vect{S};\vect{S}^{(t)})$ is
	$ \widetilde{h} (\vect{S};\vect{S}^{(t)}) = \widetilde{f}_D (\bD;\vect{S}^{(t)}) + \widetilde{f}_Z (\bZ;\vect{S}^{(t)}) + \lambda \norm{\bZ}_1 $
and the approximate problem reads
\begin{equation}
	\label{prob:PRDLapprox2}
	(\widetilde{\bD}^{(t)}, \widetilde{\bZ}^{(t)}) = \underset{\bD \in \set{D},\bZ}{\arg \min}  \ \widetilde{h} (\vect{S};\vect{S}^{(t)}).
\end{equation}
The columns of $\bD$ and all the entries of $\bZ$ are separable in the objective function of~\eqref{prob:PRDLapprox2} and the constraint set $\set{D}$ is a Cartesian product of compact convex sets, each of which involves one column $\bd_p$. Consequently, problem~\eqref{prob:PRDLapprox2} can be decomposed into $P + (P \times I)$ subproblems. Each subproblem exclusively depends on a column $\bd_p$ or a single variable $z_{p,i}$ and, hence, can be solved in parallel.

Define $\Delta \bD \! =\! \widetilde{\bD}^{(t)} \!-\! \bD^{(t)}$ and $\Delta \bZ \!=\! \widetilde{\bZ}^{(t)} \!-\! \bZ^{(t)}$. According to~\cite[Prop. 1]{yangUnifiedSuccessivePseudoconvex2017}, the difference $(\Delta \bD,\Delta \bZ)$ is a descent direction of the majorizing function $\widehat{h}(\vect{S};\vect{S}^{(t)})$ in the domain of~\eqref{prob:PRDLMatrixForm}. Thus, the following simultaneous update rule can be applied:
\begin{equation}
	\label{eq:update}
	\bD^{(t+1)} \!=\! \bD^{(t)} \!+\! \gamma^{(t)} \Delta \bD \quad \text{and} \quad \bZ^{(t+1)} \!=\! \bZ^{(t)} \!+\! \gamma^{(t)} \Delta \bZ,
\end{equation}
where $\gamma^{(t)} \in [0,1]$ is the step size. When $(\widetilde{\vect{D}}^{(t)}, \widetilde{\vect{Z}}^{(t)}) = (\vect{D}^{(t)},\vect{Z}^{(t)})$, a stationary point, in fact, a global minimizer, of $\widetilde{h} (\vect{S};\vect{S}^{(t)})$ is achieved, which is also stationary for the majorizing problem and the original problem~\eqref{prob:PRDLMatrixForm} (see Appendix~\ref{appendix:convergence}).



In the following, we describe the efficient solution approaches for the subproblems decomposed from~\eqref{prob:PRDLapprox2}. 

\textbf{Descent direction for $ \bD $.}
The $P$ independent subproblems decomposed from problem~\eqref{prob:PRDLapprox2} involving $\bD$ can be written as
\begin{equation}
	\label{prob:approx_dp}
	\begin{matrix}
	\min_{\bd_p} \ \widehat{f} ( \bd_p,\bD^{(t)}_{-p},\bZ^{(t)}; \vect{S}^{(t)} ) \quad 
	\text{s.t.} \ \tfrac{1}{2} \big(\norm{\bd_p}_2^2 - 1\big) \leq 0.
	\end{matrix}
\end{equation}
Each subproblem in~\eqref{prob:approx_dp} is an $ \ell_2 $-norm constrained LS, which has no closed-form solution.
However, as Slater's condition is satisfied for~\eqref{prob:approx_dp}, strong duality holds and, hence, the primal and dual optimal solutions can be obtained by solving the Karush-Kuhn-Tucker (KKT)
optimality system~\cite[Sec. 5.5.3]{boydConvexOptimization2004}.
By vectorization, we express $\widehat{f} ( \bd_p,\bD^{(t)}_{-p},\bZ^{(t)}; \vect{S}^{(t)} )$ as
\begin{equation}
	\label{eq:approx_dp}
	\widehat{f} ( \bd_p,\bD^{(t)}_{-p},\bZ^{(t)}; \vect{S}^{(t)} ) = \tfrac{1}{2} \norm{ \vec (\bY^{(t)}_p) - \bH_p \bd_p }_2^2,
\end{equation}
where $\bY^{(t)}_p = \bY^{(t)} - \opr{F} (\bD_{-p}^{(t)} {\bZ}_{-p}^{(t)})$ with $\bZ_{-p} \in \Cbb^{(P-1) \times I}$ obtained by removing the $p$th row of $\bZ$, and $\bH_p \!=\! \bF \cdot \big( \bz_{p:}^{(t)} \otimes \bI_{N} \big) $ with $ \bF $ in~\eqref{eq:linearOpr_vectorized}.
Then the Lagrangian associated with~\eqref{prob:approx_dp} is 
	$ L ({\bd_p},\nu_p) = \tfrac{1}{2} \norm{\vec (\bY^{(t)}_p) - \bH_p \bd_p }^2_\tF + \tfrac{\nu_p}{2} ( \norm{{\bd_p}}^2_2 - 1) $,
where $\nu_p \! \geq \! 0$ is a Lagrangian multiplier. 
Let $\widetilde{\bd}_p^{(t)}$ and $\widetilde{\nu}_p^{(t)}$ be a pair of primal and dual optimal solutions,
and let $\bH_p \!=\! \bU \bSigma \bV^\tH$ be the compact singular value decomposition (SVD) of $\bH_p$ and $\sigma_1 
\! \geq \! \cdots \! \geq \! \sigma_r \!>\! 0$ the nonzero singular values with $r \!=\! \rank(\bH_p), \ \bU \in \Cbb^{M_1 M_2 \times r}, \ \bSigma \in \Cbb^{r \times r}$, and $ \bV \in \Cbb^{N \times r} $. 
The solution $\widetilde{\bd}_p^{(t)}$ of problem~\eqref{prob:approx_dp} holds
\begin{equation}
	\label{eq:d_hat}
	\widetilde{\bd}_p^{(t)}= \bV \big( \bSigma^\tH \bSigma + \widetilde{\nu}_p^{(t)} \bI_r \big)^\dagger \bSigma^\tH \bU^\tH \vec (\bY_p^{(t)})
\end{equation}
by solving the KKT system. Define the rational function
\begin{equation}
	\label{eq:rational}
	\begin{matrix}
		\psi_p (\nu_p) = \sum_{i=1}^r \frac{ \abs{c_{i,p}}^2 }{ (\sigma_i^2 + \nu_p)^2 }
	\end{matrix} \quad \text{with} \ \bc_p = \bSigma^\tH \bU^\tH \vec (\bY_p^{(t)}).
\end{equation}
The dual optimal point $\widetilde{\nu}_p^{(t)}$ required in~\eqref{eq:d_hat} is determined by
\begin{equation}
	\label{eq:nu_hat}
	\begin{cases}
		\widetilde{\nu}_p^{(t)} = 0 , &  \text{if } \psi_p(0) \leq 1,\\
		\widetilde{\nu}_p^{(t)} \in \{ \nu_p > 0 \mid \psi_p(\nu_p) = 1\}, & \text{otherwise.}
	\end{cases}
\end{equation}
In the case where $ \psi_p(0) > 1 $, $\widetilde{\nu}_p^{(t)}$ is the unique solution of
\begin{equation}
	\label{eq:rationalEq}
	 \psi_p(\nu_p) = 1 \quad \text{for} \ \nu_p \in (0, + \infty),
\end{equation}
which has no closed-form expression, except for the case where all singular values $\sigma_i$ are identical.
In the general case, to solve~\eqref{eq:rationalEq}, 
we develop an efficient iterative algorithm based on successive rational approximation (cf.~\cite{bunchRankOneModificationSymmetric1978,liSolvingSecularEquations1993}), which is outlined in Algorithm~\ref{alg:rational} and will be described in Section~\ref{subsec:rational}.

For the particular cases with linear operator $\opr{F}$ in~\eqref{eq:oprF_time-invariant} that are investigated in the simulations, the SVD of $ \bH_p $ can be calculated analytically given the SVD of $ \bA $. 
Hence, the complexity is significantly reduced compared to the general case where an iterative algorithm, e.g., QR algorithm~\cite{golubMatrixComputations2013}, is needed to obtain the SVD of $ \bH_p $ for every column $ \bd_p $ in each iteration. Then the proposed SCA algorithm for the cPRDL problem in~\eqref{prob:PRDLMatrixForm} is competitive with that for the PRDL problem in~\eqref{prob:SC-PRIME} in terms of complexity. 
Details on the simplified solution approach for $\opr{F}$ in \eqref{eq:oprF_time-invariant} can be found in Appendix~\ref{appendix:subprobD}.

\textbf{Descent direction for $ \bZ $.}
The subproblem decomposed from~\eqref{prob:PRDLapprox2} involving each entry $ z_{p,i} $ is a univariate LASSO~\cite{tibshiraniRegressionShrinkageSelection1996} in Lagrangian form, which admits a closed-form solution
\begin{equation}
	\label{eq:softThres}
	\widetilde{z}^{(t)}_{p,i}  \!=\! \tfrac{1}{\norm{\bF_i \bd_p^{(t)}}_2^2} {{\cal S}_\lambda \! \Big( \norm{\bF_i \bd_p^{(t)}}_2^2 z_{p,i}^{(t)} \!-\! \nabla_{\!\! z_{p,i}} \widehat{f} (\vect{S}^{(t)};\vect{S}^{(t)}) \Big)}.
\end{equation}
Matrix $ \bF_i $ in~\eqref{eq:softThres} is the $ i $th block of $ \bF $ in the partition
\begin{equation}
	\label{eq:partition_F}
	\bF = [\bF_1, \ldots, \bF_I] \quad \text{with } \bF_i \in \Cbb^{M_1 M_2 \times N} \text{ for } i = 1,\ldots,I.
\end{equation}

\begin{algorithm}[t]
	\caption{compact-SCAphase} 
	\label{alg1}
	\KwIn{$\bY \in \Rbb_{+}^{M_1 \times M_2}$, $\lambda \geq 0$, tolerance $\varepsilon > 0$}
	Initialize $\bD^{(0)} \in \set{D}$ and $\bZ^{(0)}$ randomly, $t \gets 0$\;
	\While{stopping criterion~\eqref{eq:stopCriterion1} not achieved}{
		{\For(in parallel){$ p = 1, \ldots,P $}{
				$ \bH_p \gets \bF \cdot \big( \bz_{p:}^{(t)} \otimes \bI_{N} \big) $\;
				Compute the compact SVD of $ \bH_p $\;
				Compute dual optimal value $ \nu_p^{(t)} $ using~\eqref{eq:nu_hat}\;
				Compute $ \widetilde{\bd}_p^{(t)} $ according to~\eqref{eq:d_hat}\;
			}
		}
		
		{\For(in parallel){$ p = 1, \ldots,P,\ i = 1, \ldots,I $}{
				Compute $ \widetilde{z}_{p,i}^{(t)} $ according to~\eqref{eq:softThres}\;
			}
		}
		
		Compute step size $\gamma^{(t)}$ by exact line search~\eqref{prob:lineSearch}\;
		Update the variables using~\eqref{eq:update} and
		$t \gets t+1$\;
		
	}
	\Return $\bD^{(t)}, \bZ^{(t)}$
\end{algorithm}

\subsection{Step Size Computation} 
\label{subsec:alg1_stepsize}

The majorizing function $\widehat{h}$ is nonsmooth due to the regularization $g$. Thus, to efficiently find a proper step size $\gamma^{(t)}$ for the update in~\eqref{eq:update}, we follow~\cite{yangUnifiedSuccessivePseudoconvex2017} and perform an exact line search on a differentiable upper bound of $\widehat{h}$.
Ignoring constants, we can write the computation of step size $\gamma^{(t)}$ as
\begin{equation}
	\label{prob:lineSearch}
	\gamma^{(t)} = \underset{0 \leq \gamma \leq 1}{\arg \min} \left\{
	\begin{array}{l}
		\widehat{f} \big(\bD^{(t)} + \gamma \Delta \bD, \bZ^{(t)} + \gamma \Delta \bZ; \vect{S}^{(t)}\big) \\
		\qquad + \gamma \big( g(\widetilde{\bZ}^{(t)}) - g(\bZ^{(t)}) \big)
	\end{array}
	\right\},
\end{equation}
which is a minimization of fourth-order polynomial on the interval $[0,1]$ and can be solved by computing the real roots of its derivative, a cubic polynomial, in $[0,1]$ with the well-known cubic formula.
If multiple roots are found in $[0,1]$, evaluating the objective function in~\eqref{prob:lineSearch} is then needed to obtain $\gamma^{(t)}$.

The line search~\eqref{prob:lineSearch} always finds a nonzero step size $\gamma^{(t)}$ since $(\Delta \bD,\Delta \bZ)$ is a descent direction of $\widehat{h}$, until a stationary point of $h$ is attained.
With the step size $\gamma^{(t)}$ obtained by the line search~\eqref{prob:lineSearch}, the update~\eqref{eq:update} then ensures a monotonic decrease of the original objective function $h$ in~\eqref{prob:PRDLMatrixForm}, cf.~\cite{yangUnifiedSuccessivePseudoconvex2017}. 

Finally, the proposed compact-SCAphase algorithm for solving the cPRDL problem in~\eqref{prob:PRDLMatrixForm} is outlined in Algorithm~\ref{alg1}.

\subsection{Rational Approximation}
\label{subsec:rational}

\begin{algorithm}[t]
	\caption{\mbox{Rational Approximation for Solving \eqref{eq:rationalEq}.}}
	\label{alg:rational}
	\KwIn{Rational function $\psi(\nu)$, tolerance $\eta>0$}
	Initialize $\nu^{(0)} \gets 0,\ l \gets 0$\;
	\While{$\psi(\nu^{(l)}) > 1 + \eta$}{
		$\nu^{(l+1)} \gets \nu^{(l)} + {2 \psi(\nu^{(l)})} \big(1- \sqrt{\psi(\nu^{(l)})}\big) \big/ {\psi'(\nu^{(l)})}$\;
		$l \gets l+1$\;
	}
	\Return $\nu^{(l)}$
	
\end{algorithm}

Borrowing the idea in~\cite{bunchRankOneModificationSymmetric1978,liSolvingSecularEquations1993}, we develop a successive rational approximation algorithm, outlined in Algorithm~\ref{alg:rational}, for efficiently solving the rational equation~\eqref{eq:rationalEq}, which yields the dual optimal solution of~\eqref{prob:approx_dp}.
We omit the column index $p$ in the derivations below as we discuss only one column. 


Let $\nu^{(l)}$ be the approximate solution at iteration $l$.
As $\psi(\nu)$ has all negative poles, it decreases monotonically in $[0,+\infty)$.
Hence, we interpolate $\psi(\nu)$ at $ \nu^{(l)} $ by a simple rational function 
\begin{equation} \label{eq:rationalSimple}
	F(\nu;\alpha,\beta) = {\alpha} / {(\beta-\nu)^2} ,
\end{equation}
where parameters $\alpha$ and $\beta$ are chosen such that
	$ F(\nu^{(l)};\alpha,\beta) = \psi(\nu^{(l)}) \text{ and } F'(\nu^{(l)};\alpha,\beta) = \psi'(\nu^{(l)}) $.
It is easily verified that 
\begin{equation}
	\label{eq:alphabeta}
	\alpha = {4 \big( \psi(\nu^{(l)})\big)^3 } \big/ {\big(\psi'(\nu^{(l)}) \big)^2}, \quad
	\beta = \nu^{(l)} + {2 \psi(\nu^{(l)})} \big/ {\psi'(\nu^{(l)})}.
\end{equation}
Then the unique solution of $F(\nu;\alpha,\beta) \!=\! 1$ in $ (0,+\infty) $ is chosen as the next iterate $\nu^{(l+1)}$.
Omitting intermediate calculations, we can express the update rule at the $l$th iteration as
\begin{equation} \label{eq:rationalUpdate}
	\begin{matrix}
		\nu^{(l+1)} = \nu^{(l)} + {2 \psi(\nu^{(l)})} \big(1- \sqrt{\psi(\nu^{(l)})}\big) \big/ {\psi'(\nu^{(l)})}
	\end{matrix}.
\end{equation}

Define
$ \delta_i \!=\! - \sigma_i^2$, $i \!=\! 1, \ldots,r $,
which are the poles of $\psi$ with $\delta_1 \! \leq \! \ldots \! \leq \! \delta_r \! <0$. Ignoring the trivial case where all poles $ \delta_i $ are identical, we derive the following bounding property.
\begin{thm} \label{thm:rationalLowerBounding}
	$ F(\nu;\alpha,\beta) < \psi (\nu) $ for all $\nu > \delta_r$ and $\nu \neq \nu^{(l)}$.
\end{thm}

\textit{Proof:} See Appendix~\ref{appendix:rationalLowerBounding}.

Thus, if $\psi(\nu^{(l)}) \!>\! 1$, i.e., $\nu^{(l)}$ is below the solution $\widetilde{\nu}$ of equation $\psi(\nu) \!=\! 1$, then the solution of $F(\nu;\alpha,\beta) \!=\! 1$ falls between $\nu^{(l)}$ and $\widetilde{\nu}$, i.e., $\nu^{(l)} \!<\! \nu^{(l+1)} \!<\! \widetilde{\nu}$. Hence, using the proposed rational approximation, we monotonically approach $\widetilde{\nu}$ from an initial point $\nu^{(0)} \!<\! \widetilde{\nu}$.
Moreover, as we solve the rational equation in the case where $\psi(0) \!>\! 1$, $\nu$ can be simply initialized as $\nu^{(0)} \!=\! 0$. 

Like Newton's method, Algorithm~\ref{alg:rational} can be shown to have an asymptotically quadratic convergence. However, whereas Newton's method successively interpolates $\psi$ by its tangent, Algorithm~\ref{alg:rational} interpolates $\psi$ by a rational function, which leads to faster convergence due to the convexity of the rational functions in the considered interval. 
In the simulations, Algorithm~\ref{alg:rational} usually attains an accuracy of $10^{-9}$ within 4 iterations.

\subsection{Stopping Criterion}
\label{subsec:alg1_stopping}

As mentioned in Section~\ref{subsec:approx_1}, if $\vect{S}^{(t)}$ is stationary for the majorizing function $\widehat{h}(\vect{S};\vect{S}^{(t)})$, it is also stationary for the original problem~\eqref{prob:PRDLMatrixForm}. Thus, to evaluate the quality of solution, we first derive the following stationarity condition for $\widehat{h}(\vect{S};\vect{S}^{(t)})$ in the domain of problem~\eqref{prob:PRDLMatrixForm}
according to the C-stationarity defined in Section~\ref{subsec:convergence}:
for all $p = 1,\ldots,P$ and $i=1,\ldots,I$,
\begin{subequations}
	\label{eq:stationaryCondition}
	\begin{gather}
		\label{eq:stationaryConditionD}
		\nabla_{\! \bd_p} \widehat{f} (\vect{S};\vect{S}^{(t)}) \! = \!
		\begin{cases}
			\!\bzero, & \norm{{\bd}_p}_2 \!<\! 1,\\
			\!-\! \norm{\nabla_{\! \bd_p} \widehat{f} (\vect{S};\vect{S}^{(t)})}_2  {\bd}_p, & \norm{{\bd}_p}_2 \!=\! 1,
		\end{cases}\\
		\label{eq:stationaryConditionZ}
		\text{and} \quad
		\begin{cases}
			\nabla_{\! z_{p,i}} \widehat{f} (\vect{S};\vect{S}^{(t)}) = -\lambda \euler^{\imagunit \arg (z_{p,i})}, & z_{p,i} \neq 0,\\
			\abs{\nabla_{\! z_{p,i}} \widehat{f} (\vect{S};\vect{S}^{(t)})} \leq \lambda, &  z_{p,i} = 0.
		\end{cases}
	\end{gather}
\end{subequations}
Then we define the minimum-norm subgradient\footnote{It is the (Clarke) subgradient with the minimum $ \ell_2 $-norm for the extended-value extension of $ \widehat{h} $, whose values at points with $ \bD \notin \set{D}$ are set to infinity.} $ \nabla^\mathsf{S} \widehat{h} $ of an extension of $\widehat{h}$ as follows~\cite{liuBlockCoordinateDescent2019}: for all $ p = 1,\ldots,P $ and $ i=1,\ldots,I $,
\begin{align*}
	&\nabla_{\! \bd_p}^\mathsf{S} \widehat{h}(\vect{S};\vect{S}^{(t)}) = \\
	&\ \begin{cases} 
		\hspace*{-2pt} {\nabla_{\! \bd_p} \widehat{f} (\vect{S};\vect{S}^{(t)})}, & \hspace*{-3pt} \norm{\bd_p}_2 \!<\! 1,\\
		\hspace*{-2pt} \nabla_{\! \bd_p} \widehat{f}(\vect{S};\vect{S}^{(t)}) - \tfrac{ \min \big\{ 0, \Re \big(\vect{d}_p^\tH \nabla_{\! \bd_p} \widehat{f}(\vect{S};\vect{S}^{(t)})\big) \big\} }{\norm{\nabla_{\! \bd_p} \widehat{f}(\vect{S};\vect{S}^{(t)})}_2} \bd_p, & \hspace*{-3pt} \norm{\bd_p}_2 \!=\! 1,
	\end{cases} \\
	&\nabla_{\! z_{p,i}}^\mathsf{S} \widehat{h}(\vect{S};\vect{S}^{(t)}) = \begin{cases}
		\nabla_{\! z_{p,i}} \widehat{f}(\vect{S};\vect{S}^{(t)}) + \lambda \euler^{\imagunit \arg (z_{p,i})}, &  z_{p,i} \neq 0,\\
		\max \{ 0, \abs{\nabla_{\! z_{p,i}} \widehat{f}(\vect{S};\vect{S}^{(t)}) } - \lambda\} , &  z_{p,i} = 0.
	\end{cases}
\end{align*}
The minimum-norm subgradient $\nabla^\mathsf{S} \widehat{h} (\vect{S};\vect{S}^{(t)})$ vanishes at $ \vect{S}^{(t)} $ if and only if $\vect{S}^{(t)}$ fulfills the stationarity conditions~\eqref{eq:stationaryCondition}. 
This leads to a termination criterion that the minimum-norm subgradient must be small, i.e., given a tolerance $ \varepsilon >0 $,
\begin{equation}
	\label{eq:stopCriterion1}
	\begin{aligned}
		&\norm{\nabla_{\! \bD}^\mathsf{S} \widehat{h} (\vect{S}^{(t)};\vect{S}^{(t)})}_\tF \leq M_1 M_2 \cdot \sqrt{NP} \cdot \varepsilon \\
		\text{and}\quad &\norm{\nabla_{\! \bZ}^\mathsf{S} \widehat{h} (\vect{S}^{(t)};\vect{S}^{(t)})}_\tF \leq M_1 M_2 \cdot \sqrt{PI} \cdot \varepsilon,
	\end{aligned}
\end{equation}
where the sizes of measurements and variables are considered.

\section{Proposed Algorithm for Formulation PRDL}
\label{sec:alg2}

With the increase of diversity of linear measurement operator $\opr{F}$, the per-iteration complexity of compact-SCAphase dramatically grows due to the computation of partial Hessians and SVD of $\bH_p$. 
Therefore, in this section, we propose the SCAphase algorithm for the conventional formulation~\eqref{prob:SC-PRIME} based on the same extended-SCA framework as in Section~\ref{sec:alg1}.

Let $h(\bX,\bD,\bZ) = f(\bX,\bD,\bZ) + g(\bZ)$ denote the objective function in~\eqref{prob:SC-PRIME} with $ f(\bX,\bD,\bZ) = \tfrac{1}{2} \norm{\bY - \abs{\opr{F}(\bX)}}^2_\tF + \tfrac{\mu}{2} \norm{\bX - \bD \bZ}^2_\tF $ and $ g(\bZ) = \rho \norm{\bZ}_1 $.
The first component $f$ is nonconvex and nonsmooth, and the sparsity regularization $g$ is convex but nonsmooth.
In each iteration, we first find a descent direction by solving a separable convex approximate problem that is constructed based on a smooth majorization of $f$. Then all variables are jointly updated along the descent direction by exact line search, which ensures a decrease of the original function~$h$.
%
An optional debiasing step, similar to that in Section~\ref{sec:alg1}, can be applied to the PRDL problem after a stationary point is obtained, to further improve the accuracy.

\subsection{Smooth Majorization and Separable Convex Approximation}

Similarly, let $ \vect{S} = (\vect{X},\vect{D},\vect{Z}) $ be the collection of all variables. 
In iteration $ t $, we first derive a smooth majorizing function
\begin{equation}
	\label{eq:SC-PRIME_major}
	\widehat{f} (\vect{S};\vect{S}^{(t)}) = \tfrac{1}{2} \norm{\bY^{(t)} - \opr{F}(\bX) }^2_\tF + \tfrac{\mu}{2} \norm{\bX - \bD \bZ}^2_\tF
\end{equation}
for $f$ at the current point $\vect{S}^{(t)} = (\bX^{(t)},\bD^{(t)},\bZ^{(t)})$ with $\bY^{(t)} = \bY \odot \euler^{\imagunit \arg(\opr{F}(\bX^{(t)}))}$ by the property in~\eqref{eq:upbound}.
It has the partial gradients
\begin{equation}
\label{eq:grad_upperBoundf_2}
\begin{aligned}
	\nabla_{\! \bX} \widehat{f} (\vect{S};\vect{S}^{(t)}) &= \opr{F}^*( \opr{F}(\bX) - \bY^{(t)} ) + \mu  (\bX - \bD\bZ), \\ 
	\nabla_{\! \bD} \widehat{f} (\vect{S};\vect{S}^{(t)}) &= \mu (\bD \bZ - \bX) \bZ^\tH, \\ 
	\nabla_{\! \bZ} \widehat{f} (\vect{S};\vect{S}^{(t)}) &= \mu \bD^\tH (\bD \bZ - \bX). 
\end{aligned}
\end{equation}
Note that $\widehat{f}$ is nonconvex due to the bilinear map $\bD \bZ$.
Then
$ \widehat{h} (\vect{S};\vect{S}^{(t)}) = \widehat{f} (\vect{S};\vect{S}^{(t)}) + g(\bZ) $
is a majorization of $h$ at $\vect{S}^{(t)}$.

We then minimize a separable convex approximation of the majorizing function $\widehat{h}$ as $\widehat{h}$ is expensive to minimize exactly. The best-response approximation for $\widehat{h}$ is given by
\begin{equation*}
	\widetilde{h} (\vect{S};\vect{S}^{(t)}) = \widetilde{f}_X (\bX;\vect{S}^{(t)}) \!+\! \widetilde{f}_D (\bD;\vect{S}^{(t)}) \!+\! \widetilde{f}_Z (\bZ;\vect{S}^{(t)}) \!+\! \rho \norm{\bZ}_1,
\end{equation*}
where $\widetilde{f}_X (\bX;\vect{S}^{(t)}),\ \widetilde{f}_D (\bD;\vect{S}^{(t)})$ and $\widetilde{f}_Z (\bZ;\vect{S}^{(t)})$ denote the approximate functions of $\widehat{f}$ over three block variables, respectively.
The approximate functions $\widetilde{f}_D$ and $\widetilde{f}_Z$ are constructed in the same way as~\eqref{eq:PRDLapprox_f} in Section~\ref{sec:alg1}.
To limit the complexity of minimizing $\widetilde{h}$, we perform the best-response approximation on each entry of  $\bX$, which leads to the approximation
\begin{equation*}
	\widetilde{f}_X(\bX;\vect{S}^{(t)}) =
		\sum_{i=1}^I \sum_{n=1}^{N} \widehat{f} (x_{n,i}, \bX^{(t)}_{-(n,i)}, \bD^{(t)}, \bZ^{(t)}; \vect{S}^{(t)}),
\end{equation*}
where $\bX_{-(n,i)}$ is the collection of all entries of $\bX$ except $x_{n,i}$.
The approximate problem at the $t$th iteration then reads
\begin{equation}
	\label{prob:SC-PRIMEapprox}
	(\widetilde{\bX}^{(t)},\widetilde{\bD}^{(t)},\widetilde{\bZ}^{(t)}) = \underset{\bX,\bD \in \set{D},\bZ}{\arg \min} \ \widetilde{h}(\vect{S};\vect{S}^{(t)}).
\end{equation}
Likewise, problem~\eqref{prob:SC-PRIMEapprox} can be decomposed into independent subproblems, each of which exclusively depends on a column $\bd_p$ or a single variable $x_{n,i}$ or $z_{p,i}$ and can be solved in parallel.


Define~$\Delta \bX = \widetilde{\bX}^{(t)} - \bD^{(t)}$, $\Delta \bD = \widetilde{\bD}^{(t)} - \bD^{(t)}$,~and~$\Delta \bZ = \widetilde{\bZ}^{(t)} - \bZ^{(t)}$. Then the following simultaneous update rule along the descent direction $(\Delta \bX,\Delta \bD,\Delta \bZ)$ of $\widehat{h}(\vect{S};\vect{S}^{(t)})$ is applied:
\begin{multline}
	\label{eq:update_2}
	(\bX^{(t+1)},\bD^{(t+1)},\bZ^{(t+1)}) \!=\! (\bX^{(t)},\bD^{(t)}, \bZ^{(t)}) \\ \!+\! \gamma^{(t)} (\Delta \! \bX,\Delta \! \bD,\Delta \! \bZ)
\end{multline}
with ${\gamma^{(t)} \in [0,1]}$ being the step size.
When $(\widetilde{\bX}^{(t)}, \widetilde{\bD}^{(t)}, \widetilde{\bZ}^{(t)}) = (\bX^{(t)},\bD^{(t)}, \bZ^{(t)})$,
the algorithm has converged to a stationary point of the convex approximation $\widetilde{h}(\vect{S};\vect{S}^{(t)})$, which is also stationary for the majorization and the original problem~\eqref{prob:PRDLMatrixForm}.

In the following, the closed-form solutions for the subproblems decomposed from~\eqref{prob:SC-PRIMEapprox} are derived.
First, since $\widehat{f}$ is quadratic with respect to $\bX$, each subproblem involving an entry $ x_{n,i} $ is a univariate quadratic program and has a solution
\begin{equation}
	\label{eq:x_ni_hat_2}
	\widetilde{x}_{n,i}^{(t)} = x_{n,i}^{(t)} - {\nabla_{\! x_{n,i}} \widehat{f} (\vect{S}^{(t)};\vect{S}^{(t)})}/{\nabla_{\! x_{n,i}}^2 \widehat{f}},
\end{equation}
where $ \nabla_{\! x_{n,i}}^2 \! \widehat{f} \!=\! \norm{\vect{f}_{n+(i-1)N}}_2^2 + \mu $ with $ \vect{f}_{n+(i-1)N} $ being the $( n+(i-1)N) $th column of $ \bF $ in~\eqref{eq:linearOpr_vectorized}.
Next, the $P$ independent subproblems decomposed from~\eqref{prob:SC-PRIMEapprox} that involve $\bD$ are
\begin{equation}
	\label{prob:approx_dp_2}
	\begin{aligned}
		\widetilde{\bd}_p^{(t)} = \underset{\bd_p}{\arg \min} & \ \tfrac{1}{2} \norm{ \bX^{(t)} - \bD_{-p}^{(t)} \bZ_{-p}^{(t)} - \bd_p {{\bz}^{(t)}_{p:}}^\tT }^2_\tF \\ 
		\st  & \ \norm{\bd_p}_2 \leq 1,
	\end{aligned}
\end{equation}
which can again be solved via the KKT optimality system.
Unlike~\eqref{prob:approx_dp}, problem~\eqref{prob:approx_dp_2} has a simple closed-form solution
\begin{equation}
	\label{eq:d_hat_2}
	\widetilde{\bd}_p^{(t)} = \tfrac{\widehat{\bd}_p}{\max \{1,\norm{\widehat{\bd}_p}_2\}} \quad \text{with }  \widehat{\bd}_p = \bd_p^{(t)} - \tfrac{\nabla_{\bd_p} \widehat{f} (\vect{S}^{(t)};\vect{S}^{(t)}) }{\mu \norm{\bz_{p:}^{(t)}}_2^2}.
\end{equation}
Then each subproblem involving an entry $z_{p,i}$ is a Lagrangian form of univariate LASSO and has a closed-form solution~\cite{donohoDenoisingSoftthresholding1995}
\begin{equation}
	\label{eq:softThres_2}
	\hspace*{-1pt} \widetilde{z}^{(t)}_{p,i}  = \tfrac{1}{\norm{\bd_{p}^{(t)}}_2^2} {\cal S}_{\frac{\rho}{\mu}} \! \Big( \norm{\bd_{p}^{(t)}}_2^2  z_{p,i}^{(t)} \!-\! \tfrac{1}{\mu}  \nabla_{\!\! z_{p,i}} \widehat{f} (\vect{S}^{(t)};\vect{S}^{(t)}) \Big).
\end{equation}

\subsection{Step Size Computation}

Similarly to Section~\ref{subsec:alg1_stepsize}, to efficiently find a step size $\gamma^{(t)}$ for the update in~\eqref{eq:update_2} that ensures a decrease of the original function in~\eqref{prob:SC-PRIME}, we perform an exact line search on a differentiable upper bound of $\widehat{h}$, which is formulated as
\begin{multline}
	\label{prob:lineSearch_2}
	\gamma^{(t)} \! =\! \underset{0 \leq \gamma \leq 1}{\arg \min} \ \Big\{
		\widehat{f} \! \big(\bX^{(t)} \!+\! \gamma \Delta \bX, \bD^{(t)} \!+\! \gamma \Delta \bD, \bZ^{(t)} \!+\! \gamma \Delta \bZ; \vect{S}^{(t)} \big) \\
		 + \gamma \big( g(\widetilde{\bZ}^{(t)}) - g(\bZ^{(t)}) \big) \Big\}.
\end{multline}
Problem~\eqref{prob:lineSearch_2} is also a minimization of fourth-order polynomial and can be solved analytically by rooting the derivative of the objective function; we omit the straightforward details.

Finally, the proposed SCAphase algorithm for solving the PRDL problem in~\eqref{prob:SC-PRIME} is outlined in Algorithm~\ref{alg2}.

\subsection{Stopping Criterion}

If $\vect{S}^{(t)}$ is stationary for the majorizing function $\widehat{h}(\vect{S};\vect{S}^{(t)})$, then it is also stationary for the original problem~\eqref{prob:SC-PRIME}. Hence, analogously to Section~\ref{subsec:alg1_stopping}, we derive the stationarity condition for $\widehat{h}(\vect{S};\vect{S}^{(t)})$ according to the concept of C-stationarity in Section~\ref{subsec:convergence}. Based on the stationarity condition, the minimum-norm subgradient of the extension of $\widehat{h}(\vect{S};\vect{S}^{(t)})$ is introduced to evaluate the quality of the current solution.

Similar to~\eqref{eq:stationaryCondition}, for a stationary point of $\widehat{h}(\vect{S};\vect{S}^{(t)})$, the gradients $\nabla_{\bD} \widehat{f}$ and $\nabla_{\bZ} \widehat{f}$ given in~\eqref{eq:grad_upperBoundf_2} must satisfy the following conditions:
for all $p = 1,\ldots,P$ and $i=1,\ldots,I$,
\begin{subequations}
	\label{eq:stationaryCondition_2}
	\begin{gather}
		\label{eq:stationaryConditionD_2}
		\nabla_{\! \bd_p} \widehat{f} \! (\vect{S};\vect{S}^{(t)}) \! = \!
		\begin{cases}
			\bzero, & \norm{{\bd}_p}_2 \!<\! 1,\\
			- \norm{\nabla_{\! \bd_p} \widehat{f} \! (\vect{S};\vect{S}^{(t)})}_2  {\bd}_p, &  \norm{{\bd}_p}_2 \!=\! 1,
		\end{cases}\\
		\label{eq:stationaryConditionZ_2}
		\text{and} \quad
		\begin{cases}
			\nabla_{\! z_{p,i}} \widehat{f} (\vect{S};\vect{S}^{(t)}) = -\rho \euler^{\imagunit \arg (z_{p,i})}, & z_{p,i} \neq 0,\\
			\abs{\nabla_{\! z_{p,i}} \widehat{f} (\vect{S};\vect{S}^{(t)})} \leq \rho, & z_{p,i} = 0.
		\end{cases}
	\end{gather}
\end{subequations}
Then the components of the minimum-norm subgradient with respect to matrices $\bZ$ and $\bD$ are defined in the same way as in Section~\ref{subsec:alg1_stopping}.
As for the gradient with respect to matrix $\bX$, stationarity simply requires the gradient $\nabla_{\bX} \widehat{f}$ to vanish, i.e.,
\begin{equation}
	\label{eq:stationaryConditionX1}
	\nabla_{\! \bX} \widehat{f} (\vect{S};\vect{S}^{(t)}) = \bzero.
\end{equation}
Thus, the component $\nabla_{\! \bX}^\mathsf{S} \widehat{h}$ of the minimum-norm subgradient with respect to $\bX$ is simply defined as the gradient $\nabla_{\! \bX} \widehat{f}$.

In summary, the stationary conditions of the majorizing function $\widehat{h}(\vect{S};\vect{S}^{(t)})$ consist of~\eqref{eq:stationaryCondition_2}-\eqref{eq:stationaryConditionX1}.
The algorithm is terminated when the minimum-norm subgradient is sufficiently small, i.e., with a given tolerance $\varepsilon \!>\! 0$,
\begin{equation}
	\label{eq:stopCriterion2}
	\begin{cases}
		\norm{\nabla_{\! \bD}^\mathsf{S} \widehat{h}(\vect{S}^{(t)};\vect{S}^{(t)})}_\tF \leq M_1 M_2 \cdot \sqrt{NP} \cdot \varepsilon, \\
		\norm{\nabla_{\! \bZ}^\mathsf{S} \widehat{h}(\vect{S}^{(t)};\vect{S}^{(t)})}_\tF \leq M_1 M_2 \cdot \sqrt{PI} \cdot \varepsilon,\\
		\norm{\nabla_{\! \bX}^\mathsf{S} \widehat{h}(\vect{S}^{(t)};\vect{S}^{(t)})}_\tF \leq M_1 M_2 \cdot \sqrt{NI} \cdot \varepsilon.
	\end{cases}
\end{equation}

\subsection{Comparison with SC-PRIME}
The proposed SCAphase algorithm and the state-of-the-art SC-PRIME~\cite{qiuUndersampledSparsePhase2017} adopt the same formulation, i.e., the PRDL problem in~\eqref{prob:SC-PRIME}, and the same successive majorization technique~\eqref{eq:SC-PRIME_major}. However, there are two important differences between the two algorithms.
First, SC-PRIME updates the variables in a block coordinate descent (BCD) manner, i.e., minimizes the majorizing function $\widehat{h}$ alternatively with respect to each block variable $\bX$, $\bZ$, and each column of $\bD$, instead of using parallel updates. 
Then, to avoid the expensive exact minimization of $\widehat{h}$, SC-PRIME minimizes a different separable convex approximation for each block variable from SCAphase.
Instead of using the best-response approximation, SC-PRIME further majorizes the LS function $\widehat{f}$ by replacing the partial Hessian with respect to a block variable by the identity matrix scaled by an upper bound of its eigenvalues. This majorization can be minimized in closed form and a decrease of the original objective function $h$ is ensured without a step size search. 
However, since the Hessian is typically ill conditioned, this majorization tends to be conservative, which may lead to slow convergence. In contrast, the best-response approximation $\widetilde{f}$ equivalently preserves all diagonal entries of the Hessian but is not necessarily a majorization of the original function $f$. Thus, discarding the global upper bound constraint provides more flexibility in designing an approximation that yields faster convergence to a good stationary point.
This advantage is demonstrated numerically in Section~\ref{sec:results}. 

\begin{algorithm}[t]
	\caption{SCAphase} 
	\label{alg2}
	\KwIn{ $\bY \in \Rbb_{+}^{M_1 \times M_2},\ \mu \geq 0,\ \lambda \geq 0$, tolerance $ \varepsilon > 0$}
	Initialize $\bX^{(0)} $ and $\bD^{(0)} \in \set{D}$ randomly, $\bZ^{(0)} \gets (\bD^{(0)})^\dagger \bX^{(0)}$, $t \gets 0$\;
	\While{stopping criterion~\eqref{eq:stopCriterion2} not achieved}{
		{\For(in parallel){$ n = 1, \ldots,N, \ i = 1, \ldots,I $}{
				Compute $ \widetilde{x}_{n,i}^{(t)} $ according to~\eqref{eq:x_ni_hat_2}\;
			}
		}
		{\For(in parallel){$ p = 1, \ldots,P $}{
				Compute $ \widetilde{\bd}_p^{(t)} $ according to~\eqref{eq:d_hat_2}\;
			}
		}
		
		{\For(in parallel){$ p = 1, \ldots,P, \ i = 1, \ldots,I $}{
				Compute $ \widetilde{z}_{p,i}^{(t)} $ according to~\eqref{eq:softThres_2}\;
			}
		}
		
			Compute step size $\gamma^{(t)}$ by exact line search~\eqref{prob:lineSearch_2}\;
			Update the variables using~\eqref{eq:update_2} and
		$t \gets t+1$\;
	}
	\Return $\bX^{(t)}, \bD^{(t)}, \bZ^{(t)}$
\end{algorithm}

\section{Convergence and Complexity}
\label{sec:convergence&complexity}

\begin{table*}[t]
	\centering
	\caption{Computational complexity of dominant operations in each iteration} \label{table:per-iterationComplexity}
	\begin{tabular}{|m{2.5cm}|m{3cm}|m{6cm}|m{4cm}|}
		\hline
		& computation of gradient & computation of partial Hessians & computation of polynomial coefficients in line search function \\
		\hline
		compact-SCAphase & $ c(\opr{F}) + 4NPI $ & $ \begin{matrix*}[l]
			\text{general case: } 4M_1 M_2 NPI + \mathcal{O} (M_1 M_2 N^2 P);\\ \text{special case with } \opr{F} \text{ in~\eqref{eq:oprF_time-invariant}: } 2M_1NP + 2M_2 P I
		\end{matrix*} $ 
		& $ 2c(\opr{F}) + 6NPI $ \\
		\hline
		SCAphase & $ c(\opr{F}) + 4NPI $ & $ 2NP + 2PI$ & $ c(\opr{F}) +  6NPI $\\
		\hline
		SC-PRIME & $ 2c(\opr{F}) + 6NPI $ &  --  & -- \\
		\hline
	\end{tabular}
\end{table*}

\subsection{Convergence Analysis}
\label{subsec:convergence}
For a nonsmooth optimization, the gradient consistency condition~\cite[A2.2]{sunMajorizationMinimizationAlgorithmsSignal2017} in the classic MM algorithms and BSUM requires the consistency of directional derivatives between the original nonsmooth function and its majorant at the current point in all directions, which apparently cannot be satisfied at a non-differentiable point of the original function if the majorant is restricted to be smooth, such as in the proposed algorithms. On the other hand, the convergence in the SCA framework is only established for composite problems with smooth loss functions. Hence, neither the convergence analysis of MM nor that of SCA can be applied to the proposed algorithms. In this subsection, we establish the convergence of our proposed algorithms based on a generalized concept of stationarity. 


To this end, we first introduce a generalization of the subdifferential of a function, since the usual convex subdifferential does not exist at every point for a nonconvex function.
Consider a general continuous but nonconvex and nonsmooth function $f(\bs): \Rbb^n \! \rightarrow \! \Rbb$ that is locally Lipschitz~\cite[Def. 1]{liUnderstandingNotionsStationarity2020}.
It implies that the Clarke directional derivative of $f$ exists at every point $\bs \in \Rbb^n$ in any direction $\br \in \Rbb^n$ and is defined as~\cite{clarkeOptimizationNonsmoothAnalysis1990}
\begin{equation}\label{eq:C-direcDeriv}
	f^\circ (\bs;\br) \define {\limsup}_{\substack{\bs' \rightarrow \bs, t \downarrow 0}} \ \tfrac{f(\bs' + t\br) - f(\bs')}{t}.
\end{equation}
The Clarke subdifferential (C-subdifferential) of $f$ at $\bs$ is then defined based on the Clarke directional derivative as~\cite{clarkeOptimizationNonsmoothAnalysis1990}
\begin{equation}\label{eq:C-subdiff}
	\partial_C f(\bs) \define \big\{ \bv \in \Rbb^n \mid \bv^\tT \br \leq f^\circ (\bs;\br) \text{ for all } \br \in \Rbb^n \big\}.
\end{equation}

Now consider a general constrained problem
\begin{equation}\label{prob:generalConstrained}
	\begin{matrix}
		{\min}_{\bs \in \set{C}} \ f(\bs),
	\end{matrix}
\end{equation}
where $f$ is locally Lipschitz and $\set{C} \! \subseteq \! \Rbb^n$ is a closed convex set. 
One possible generalization of stationarity for the constrained nonsmooth problem~\eqref{prob:generalConstrained} is the \textit{Clarke stationarity} (C-stationarity)~\cite{pangComputingBStationaryPoints2017,clarkeOptimizationNonsmoothAnalysis1990}, which is defined as follows.
\begin{defi} \label{defi:C-stationarity}
	(C-stationarity). A point $\bs \! \in \! \Rbb^n$ is said to be a C-stationary point of problem~\eqref{prob:generalConstrained} if it satisfies
	\begin{equation}\label{eq:C-stationarity}
		\bzero \in \partial_C f(\bs) + \set{N}_{\set{C}} (\bs), 
	\end{equation}
	where $\set{N}_{\set{C}} (\bs)$ is the Clarke normal cone of set $\set{C}$ at $\bs$~\cite{clarkeOptimizationNonsmoothAnalysis1990}.
\end{defi}

Definition~\ref{defi:C-stationarity} is motivated by the following two facts. First, condition~\eqref{eq:C-stationarity} is a necessary condition for $\bs$ being a locally minimal point of problem~\eqref{prob:generalConstrained}~\cite[Prop. 2.4.3]{clarkeOptimizationNonsmoothAnalysis1990}, but not sufficient unless problem~\eqref{prob:generalConstrained} is convex, which is similar to the usual stationarity condition in the smooth case. Second, Definition~\ref{defi:C-stationarity} is consistent with the usual concept of stationarity in the special cases where the problem is smooth or convex.
Particularly, the stationarity conditions in~\eqref{eq:stationaryCondition} and~\eqref{eq:stationaryCondition_2} for the majorization in compact-SCAphase and SCAphase, respectively, are special cases of condition~\eqref{eq:C-stationarity}.
Although the above definitions of subdifferential and stationarity are described for a problem with real-valued variables, the same concepts can be immediately extended to the complex-valued case.

Then we claim that the compact-SCAphase algorithm converges according to the following theorem.
\begin{thm} \label{prop:convergence}
	Every limit point of the solution sequence $ (\bD^{(t)},\bZ^{(t)})_t $ generated by the compact-SCAphase algorithm is a C-stationary point of problem~\eqref{prob:PRDLMatrixForm}.
\end{thm}

\textit{Proof:} See Appendix~\ref{appendix:convergence}.

A similar theorem can be claimed for SCAphase, since compact-SCAphase and SCAphase can be viewed as instances of the same extended SCA framework on different problems.

The classic MM algorithms possess the convergence to the set of directional stationary points~\cite{sunMajorizationMinimizationAlgorithmsSignal2017}, which can be shown to be a subset of C-stationary points by the definition of Clarke directional derivative.
Hence, compared to the classic MM algorithms, which, in our problem, require a nonsmooth upper bound and high computational complexity, the proposed algorithms basically sacrifice the strictness of stationarity so as to construct a surrogate problem that can be easily addressed.

In~\cite{qiuUndersampledSparsePhase2017}, the authors address the convergence of SC-PRIME, which employs the same smooth majorization in~\eqref{eq:upbound}, under the framework of BSUM~\cite{razaviyaynUnifiedConvergenceAnalysis2013}. However, the convergence analysis in~\cite{qiuUndersampledSparsePhase2017} is incomplete since the authors ignored the aforementioned fact that the gradient consistency condition required by BSUM cannot be satisfied at non-differentiable points of the original function. The convergence analysis in Appendix~\ref{appendix:convergence} can be used to fill this gap and justify that SC-PRIME converges to a stationary point of~\eqref{prob:SC-PRIME} corresponding to the same generalized concept of stationarity, i.e., C-stationarity.

In addition, another extension of SCA framework is proposed in~\cite{yangSuccessiveConvexApproximation2018} based on the difference of convex technique, which differs from our proposed algorithm in the following two aspects. First, the algorithm in~\cite{yangSuccessiveConvexApproximation2018} tackles a composite problem with a smooth but not necessarily convex loss function and a nonconvex nonsmooth regularization, whereas in this paper, as shown in~\eqref{prob:PRDLMatrixForm} and~\eqref{prob:SC-PRIME}, a composite problem with a nonconvex nonsmooth loss function and a convex but not necessarily smooth regularization is addressed. Second, a different generalization of stationarity is employed in~\cite{yangSuccessiveConvexApproximation2018} to establish the convergence.
The set of C-stationary points can be shown to be a subset of the stationary points defined in~\cite{yangSuccessiveConvexApproximation2018} by the subadditivity of C-subdifferential~\cite[Prop. 2.3.3]{clarkeOptimizationNonsmoothAnalysis1990}.

\subsection{Computational Complexity}
\label{subsec:complexity}

In this subsection, we present a theoretic comparison on the complexity of the proposed algorithms, compact-SCAphase and SCAphase, and the state-of-the-art SC-PRIME~\cite{qiuUndersampledSparsePhase2017}.

As presented in Table~\ref{table:per-iterationComplexity}, for each algorithm, we count the number of flops~\cite{golubMatrixComputations2013} required by the dominant operations, such as matrix-matrix multiplication, in each iteration, which reflects the per-iteration complexity in the worst case where the flops are executed in sequence.
The per-iteration complexity of the proposed algorithms are dominated by three components: the computation of gradient and partial Hessians of the smooth majorization $\widehat{f}$, which are required for solving the convex subproblems, and the computation of polynomial coefficients of the line search function in the step size computation. 
In the simulations, the rational approximation algorithm employed by compact-SCAphase for solving the subproblems requires 3 or 4 iterations to achieve a precision of $10^{-9}$. Therefore, the complexity of the rational approximation is comparable to that of computing a closed-form solution as in SCAphase and SC-PRIME, which is negligible compared to the other operations.
In Table~\ref{table:per-iterationComplexity}, $c(\opr{F})$ stands for the complexity of the linear operator $\opr{F}$ or, equivalently, that of its adjoint $\opr{F}^*$, which depends on the structure of $\opr{F}$ and the specific implementation. In principle, $ c(\opr{F}) $ admits the bounds 
	$ 2 NI \cdot \max \{M_1,M_2\} \leq c(\opr{F}) \leq 2M_1M_2NI $. 

Compared to SCAphase, in the general case, compact-SCAphase has a per-iteration complexity of higher order due to the computation of partial Hessians and SVD of matrix $\bH_p$ in~\eqref{eq:approx_dp}.
However, in the special case with the linear operator $\opr{F}$ in~\eqref{eq:oprF_time-invariant},
such as Cases 1 and 2 in the simulations, the complexity of computation of partial Hessians in compact-SCAphase dramatically decreases and
the SVD of $\bH_p$ can be analytically calculated given the SVD of $\bA$. Then compact-SCAphase and SCAphase have comparable per-iteration complexity. On the other hand, as shown in Fig.~\ref{fig:accuracy_varyL}, compared to SCAphase, compact-SCAphase typically uses half the number of iterations to achieve a stationary point due to the reduction of variables, which makes compact-SCAphase more competitive than SCAphase in the case with $\opr{F}$ in~\eqref{eq:oprF_time-invariant}.

Next, we compare the complexity of SCAphase and SC-PRIME. The line search is not required in SC-PRIME as it employs the BCD update. In the specific implementation of SC-PRIME used in this paper, constant rough upper bounds for the eigenvalues of the partial Hessians are used to construct the surrogate subproblems and, hence, only the gradient of $\widehat{f}$ is needed. 
However, compared to SC-PRIME, the additional line search in SCAphase does not cause a significant increase on the overall per-iteration complexity as several intermediate variables in the computation of gradient can be updated recursively. For example, $ \opr{F}(\bX^{(t)}) $ required in~\eqref{eq:grad_upperBoundf_2} is updated recursively by
	$ \opr{F}(\bX^{(t+1)}) = \opr{F}(\bX^{(t)}) + \gamma^{(t)} \opr{F} (\Delta \bX) $,
where $ \opr{F} (\Delta \bX) $ was previously calculated in the computation of coefficients of line search function.
Thus, SCAphase and SC-PRIME also have similar per-iteration complexity, especially in the case with a highly diverse linear operator $ \opr{F} $, where the per-iteration complexity is dominated by the complexity of $ \opr{F} $. 
On the other hand, with the additional line search, SCAphase exhibits faster convergence in terms of number of iterations.

Finally, we remark that, in contrast to the BCD update in SC-PRIME, the computation of solutions of subproblems in compact-SCAphase and SCAphase can be fully parallelized with suitable hardware architectures.

\section{Simulation Results}
\label{sec:results}

In this section, we compare the performance of the two proposed algorithms and the state-of-the-art SC-PRIME~\cite{qiuUndersampledSparsePhase2017} on synthetic data in the context of blind channel estimation in a multi-antenna random access network.
All experiments were conducted on a Linux machine assigned with two 2.3~GHz cores and 7~GB RAM running MATLAB R2021b.
Although, theoretically, all the subproblems in each iteration in the proposed algorithms can be solved in parallel, for simplicity, the subproblems involving different block variables (i.e., $\bX$, $\bD$, or $\bZ$) are solved sequentially, whereas the computation of solutions for subproblems involving the same block variable are parallelized by using vectorization in MATLAB.

\subsection{Simulation Setup}

\begin{figure}[t]
	\centering
	\includegraphics[width=0.95\linewidth]{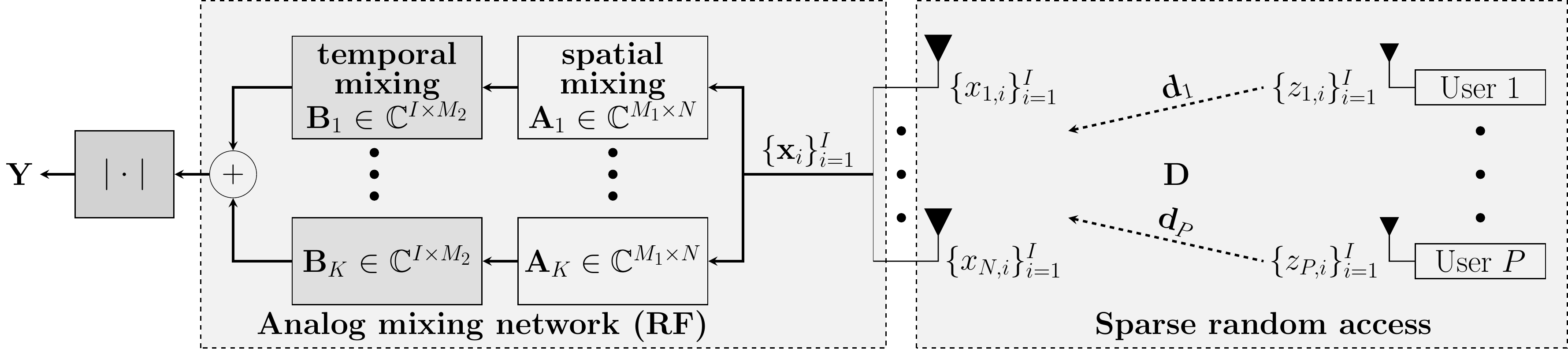}
	\caption{Multi-antenna Random Access Network.}
	\label{fig:RAN}
\end{figure}

We consider a multi-antenna random access network with magnitude-only measurements in Fig.~\ref{fig:RAN}. 
The base station is equipped with $N$ antennas and $P$ single-antenna users with unknown spatial signatures $\{ \mathbf{d}_p \! \in \! \Cbb^N \}_{p=1}^P$ sporadically access the channel in $I$ time-slots. In time-slot $i$ user $p$ transmits an unknown information symbol ${z_{p,i} \! \neq \! 0}$ with probability $L/P$ and ${z_{p,i} \! = \! 0}$ with probability ${(P-L)/P}$, where $L$ defines the expected sparsity level of the transmitted symbol vectors ${\bz}_i \!=\!  [z_{1,i}, \ldots, z_{P,i}]^\tT$, $i = 1, \ldots,I$. With ${\bD} \! = \! [{\bd}_1,\ldots, {\bd}_P]$, the received symbol vector ${\bx}_i \!=\! [x_{1,i}, \ldots, x_{N,i}]^\tT$ at the antennas is given by
$ \bx_i \! = \! \bD \bz_i, $
which cannot be directly observed due to heavy phase errors caused by the phase noise of the local oscillators in the down-converters and analog-to-digital converters~\cite{aminuBeamformingTransceiverOptimization2019}.
Hence, before down-converted and sampled, the received signals are first processed by an analog mixing network at radio frequency composed of analog phase shifters and analog filters.
Then the objective is to jointly learn the spatial signature matrix $\bD$ and the sparse transmitted symbol vectors $\bz_i$ from spatially and temporally filtered subband magnitude measurements,
which can be expressed by the model in~\eqref{eq:measurements}, whereas the heavily corrupted phase measurements are discarded.
Furthermore, the subband measurements can be acquired at a reduced sampling rate according to the bandwidth of the respective subband filters.
In this application, as shown in Fig.~\ref{fig:RAN}, the linear operator $\opr{F}$ in~\eqref{eq:oprF_general} is interpreted as $K$ independent chains of linear spatial mixing networks $\{\bA_k \! \in \! \Cbb^{M_1 \! \times \! N}\}_{k=1}^K$ and temporal mixing networks $\{\bB_k \! \in \! \Cbb^{I \! \times \! M_2}\}_{k=1}^K$.
Note that the order of the spatial and temporal mixing is interchangeable for each chain.
In our simulations, the spatial mixing networks $\{\bA_k\}_{k=1}^K$ are generated from a standard complex Gaussian distribution, and the following three particular cases of linear operator $\opr{F}$ of different levels of diversity are investigated:
\begin{itemize}
	\item \textit{Case 1: Time-invariant spatial mixing and no temporal mixing.} In this case, $\opr{F}$ is interpreted by a single chain of mixing networks, i.e., $K \!=\! 1$. For simplicity, we omit the subscript on the mixing networks and $\opr{F}$ reduces to
	\begin{equation}
		\label{eq:oprF_time-invariant}
		\opr{F}(\bX) = \bA \bX \bB.
	\end{equation}
	Moreover, the temporal mixing is set to be $\bB = \bI$.
	
	\item \textit{Case 2: Time-invariant spatial mixing and STFT temporal mixing.}
	In this case, $\opr{F}$ can also be expressed by the model~\eqref{eq:oprF_time-invariant}, whereas the temporal mixing $\bB$ is designed to be the short-time Fourier transform (STFT)~\cite{jaganathanSTFTPhaseRetrieval2016}, which can be implemented by analog subband filters.
	
	\item \textit{Case 3: Time-variant spatial mixing and no temporal mixing.}
	In this case, $\opr{F}$ is expressed by the model~\eqref{eq:oprF_general} with $K=I$, and the $k$th temporal mixing is set to be $\bB_k = [ \bzero, \ldots, \bzero, \be_k, \bzero, \ldots,\bzero]$ with $\be_k$ being a standard basis vector, which simply selects the $k$th snapshot. Also, $\bA_k$ is the spatial mixing network designed for the $k$th snapshot.
\end{itemize}

The basic simulation setup is as follows. 
In each time-slot~$i$, $L$ randomly selected elements of the true transmitted sparse signal~$\bz_i^{\text{true}}$ are set to be nonzero. The nonzero elements of matrix~$\bZ^{\text{true}}$, all elements of spatial mixing matrices~$\{\bA_k\}_{k=1}^K$ and the true spatial signature~$\bD^{\text{true}}$ are drawn from an i.i.d. standard  complex Gaussian distribution. 
The magnitude measurements~$\bY$ are generated according to~\eqref{eq:measurements} with additive white Gaussian noise. 
The number of Monte-Carlo runs is 50.

From the solutions $\bD$ and $\bZ$ obtained by compact-SCAphase, the variable $\bX$ is constructed as $\bX \! = \! \bD \bZ$ for the performance evaluation. 
Note that the analog mixing network architecture in Fig.~\ref{fig:RAN} is also applicable in other applications of the phase retrieval with dictionary learning problem such as diffraction imaging, where various optical masks and filters can be used to increase the diversity of the intensity measurements with the objective to improve the signal recovery. In this application, the signal $\bX$ is the parameter of interest, and the dictionary $\bD$ and sparse code matrix $\bZ$ are considered as nuisance parameters. 
In contrast, in the considered application of multi-antenna network, our main target is the spatial signature matrix $\bD$ and transmitted signals $\bZ$.
Hence, only the estimation qualities of $\bD$ and $\bZ$ are presented in the following simulations.
However, the solution $\bX$ is still required in the disambiguation step, which is described afterwards.

In both formulations~\eqref{prob:PRDLMatrixForm} and~\eqref{prob:SC-PRIME}, the variables can only be recovered up to three trivial ambiguities.
Specifically, any combination of the following three trivial operations conserve the magnitude measurements and the sparsity pattern of $\bZ$:
1) global phase shift: $(\bX,\bZ) \! \rightarrow \! (\bX \euler^{\imagunit\phi}, \bZ  \euler^{\imagunit\phi})$, 2) scaling: $(\bd_p,\bz_{p:}) \! \rightarrow \! (\alpha_p  \bd_p, \alpha_p^{-1}  \bz_{p:})$ with any $\alpha_p \! \in \! \Cbb$ and ${\alpha_p \! \neq \! 0}$, 3) permutation: $(\bD,\bZ) \! \rightarrow \! (\bD \bP^\tT, \bP \bZ)$ with any permutation matrix ${\bP \! \in \! \Rbb^{P \times P}}$.
Also, if no temporal mixing is applied, the signal in each time-slot is measured independently and, hence, the global phase ambiguity holds columnwise, i.e., $(\bx_i,\bz_i) \! \rightarrow \! (\bx_i \euler^{\imagunit\phi_i}, \bz_i  \euler^{\imagunit\phi_i})$.

A disambiguation step is required to measure the estimation quality of the solutions.
Let $\bX^{\text{true}} \! = \! \bD^{\text{true}} \bZ^{\text{true}}$ be the true received signals.
To resolve the global phase ambiguity, the solution $\bX$ is corrected by the global phase shift
	$ \phi^\star \!=\! {\arg \min}_{ \phi \in [0,2\pi) } { \norm{\bX \euler^{\imagunit\phi} - \bX^\text{true} }_\tF^2 } $
in the case with temporal mixing, and the phase correction is applied columnwise with
	$ \phi_i^\star \!=\! {\arg \min}_{\phi_i \in [0,2\pi)} \norm{\bx_i \euler^{\imagunit\phi_i} - \bx_{i}^{\text{true}}}^2_2 $ for $ i=1,\ldots,I $
in the case without temporal mixing.
For the permutation ambiguity on $\bD$ and $\bZ$, a heuristic method is used to find the permutation that best matches the ground-truth with respect to the normalized cross correlation between columns in $\bD$ and $\bD^{\text{true}}$. After permutation, the estimation quality of $\bD$ is evaluated by the minimum normalized squared error (MNSE) defined as 
	$ \text{MNSE}(\bD) \!=\! {\min}_{\{\alpha_p \in \Cbb\}_{p=1}^P} \! { \big(\sum_{p=1}^P \norm{\alpha_p \bd_p \!-\! \bd_p^{\text{true}}}^2_2\big) }/{\norm{\bD^{\text{true}}}_\tF^2} $.
As for $\bZ$, after permutation, we first perform the same global phase shift $\euler^{\imagunit\phi^\star}$ on $\bZ$ or $\euler^{\imagunit\phi_i^\star}$ on each column $\bz_i$ and then the MNSE of $\bZ$ is analogously calculated as 
	$ \text{MNSE}(\bZ) \!=\! {\min}_{\{\beta_p \in \Cbb\}_{p=1}^P} \! { \big(\sum_{p=1}^P \norm{\beta_p \bz_{p:} \!-\! \bz_{p:}^{\text{true}}}_2^2\big) }/{\norm{\bZ^{\text{true}}}_\tF^2} $.
Moreover, the accuracy of the support of the estimated $\bZ$ is evaluated by
	$ \text{F-measure} = {2 \text{TP}}/{(2\text{TP} + \text{FP} + \text{FN})} $,
defined from the number of correctly and incorrectly estimated nonzeros: true positives (TP), false positives (FP), and false negatives (FN)~\cite{manningIntroductionInformationRetrieval2008}.

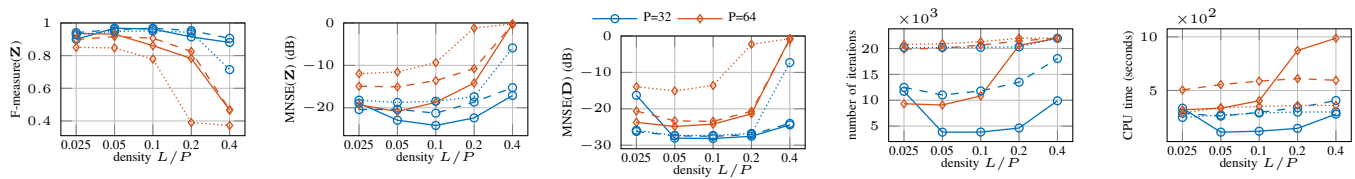
\begin{figure*}[t]
	\centering
	
	\begin{subfigure}{0.18\linewidth}
		\centering
%
%

\begin{tikzpicture}[inner ysep=1.5pt]

\begin{axis}[%
at={(0,0)},
scale only axis,
xlabel style={font=\color{white!15!black}, font=\tiny},
xlabel={density $ L/P $},
xlabel shift = -2.5pt,
xmode = log,
xtick = {0.025,0.05,0.1,0.2,0.4},
xticklabels = {0.025,0.05,0.1,0.2,0.4},
ymax=1,
ylabel style={font=\color{white!15!black}, font=\tiny},
ylabel={F-measure($\bZ$)},
ylabel shift = -3pt,
axis background/.style={fill=white},
xmajorgrids,
ymajorgrids,
legend style={legend cell align=left, align=left, draw=white!15!black, 
	font=\tiny, row sep=-2pt, draw=none,
	at={(0,0)}, anchor = south west},
cycle multi list = {
	solid,dashed,densely dotted\nextlist
	[2 of]MATLABcolormarklist
},
legend to name=legend2,
legend columns=2,
]

\addplot table [x=density, y={P=32}] {results/gauss_spatial_no_temporal/mediumScaleTest/varyL/FmeasureZ_FUN_PRDLsca.csv};
\addplot table [x=density, y={P=64}] {results/gauss_spatial_no_temporal/mediumScaleTest/varyL/FmeasureZ_FUN_PRDLsca.csv};
\addplot table [x=density, y={P=32}] {results/gauss_spatial_no_temporal/mediumScaleTest/varyL/FmeasureZ_FUN_PRDLscaX.csv};
\addplot table [x=density, y={P=64}] {results/gauss_spatial_no_temporal/mediumScaleTest/varyL/FmeasureZ_FUN_PRDLscaX.csv};
\addplot table [x=density, y={P=32}] {results/gauss_spatial_no_temporal/mediumScaleTest/varyL/FmeasureZ_scPRIME.csv};
\addplot table [x=density, y={P=64}] {results/gauss_spatial_no_temporal/mediumScaleTest/varyL/FmeasureZ_scPRIME.csv};
\legend{P=32,P=64};

\end{axis}
\end{tikzpicture}%
	\end{subfigure}
	\hfill
	\begin{subfigure}{0.18\linewidth}
		\centering
%
%

\begin{tikzpicture}[inner ysep=1.5pt]

\begin{axis}[%
at={(0,0)},
scale only axis,
xlabel style={font=\color{white!15!black}, font=\tiny},
xlabel={density $ L/P $},
xlabel shift = -2.5pt,
xmode = log,
xtick = {0.025,0.05,0.1,0.2,0.4},
xticklabels = {0.025,0.05,0.1,0.2,0.4},
ymax=0,
ylabel style={font=\color{white!15!black}, font=\tiny},
ylabel={MNSE($\bZ$) (dB)},
ylabel shift = -3pt,
axis background/.style={fill=white},
xmajorgrids,
ymajorgrids,
legend style={legend cell align=left, align=left, draw=white!15!black, 
	font=\tiny, row sep=-2pt, draw=none,
	at={(0,0)}, anchor = south west},
cycle multi list = {
	solid,dashed,densely dotted\nextlist
	[2 of]MATLABcolormarklist
},
]

\addplot table [x=density, y={P=32}] {results/gauss_spatial_no_temporal/mediumScaleTest/varyL/errorZ_FUN_PRDLsca.csv};
\addplot table [x=density, y={P=64}] {results/gauss_spatial_no_temporal/mediumScaleTest/varyL/errorZ_FUN_PRDLsca.csv};
\addplot table [x=density, y={P=32}] {results/gauss_spatial_no_temporal/mediumScaleTest/varyL/errorZ_FUN_PRDLscaX.csv};
\addplot table [x=density, y={P=64}] {results/gauss_spatial_no_temporal/mediumScaleTest/varyL/errorZ_FUN_PRDLscaX.csv};
\addplot table [x=density, y={P=32}] {results/gauss_spatial_no_temporal/mediumScaleTest/varyL/errorZ_scPRIME.csv};
\addplot table [x=density, y={P=64}] {results/gauss_spatial_no_temporal/mediumScaleTest/varyL/errorZ_scPRIME.csv};

\end{axis}
\end{tikzpicture}%
	\end{subfigure}
	\hfill
	\begin{subfigure}{0.18\linewidth}
		\centering
		\ref{legend2}
		\vspace*{-3pt}
%
%

\begin{tikzpicture}[inner ysep=1.5pt]

\begin{axis}[%
at={(0,0)},
scale only axis,
xlabel style={font=\color{white!15!black}, font=\tiny},
xlabel={density $ L/P $},
xlabel shift = -2.5pt,
xmode = log,
xtick = {0.025,0.05,0.1,0.2,0.4},
xticklabels = {0.025,0.05,0.1,0.2,0.4},
ymax=0,
ylabel style={font=\color{white!15!black}, font=\tiny},
ylabel={MNSE($\bD$) (dB)},
ylabel shift = -3pt,
axis background/.style={fill=white},
xmajorgrids,
ymajorgrids,
legend style={legend cell align=left, align=left, draw=white!15!black, 
	font=\tiny, row sep=-2pt, draw=none,
	at={(0,0)}, anchor = south west},
cycle multi list = {
	solid,dashed,densely dotted\nextlist
	[2 of]MATLABcolormarklist
},
]

\addplot table [x=density, y={P=32}] {results/gauss_spatial_no_temporal/mediumScaleTest/varyL/errorD_FUN_PRDLsca.csv};
\addplot table [x=density, y={P=64}] {results/gauss_spatial_no_temporal/mediumScaleTest/varyL/errorD_FUN_PRDLsca.csv};
\addplot table [x=density, y={P=32}] {results/gauss_spatial_no_temporal/mediumScaleTest/varyL/errorD_FUN_PRDLscaX.csv};
\addplot table [x=density, y={P=64}] {results/gauss_spatial_no_temporal/mediumScaleTest/varyL/errorD_FUN_PRDLscaX.csv};
\addplot table [x=density, y={P=32}] {results/gauss_spatial_no_temporal/mediumScaleTest/varyL/errorD_scPRIME.csv};
\addplot table [x=density, y={P=64}] {results/gauss_spatial_no_temporal/mediumScaleTest/varyL/errorD_scPRIME.csv};

\end{axis}
\end{tikzpicture}%
	\end{subfigure}
	\hfill
	\begin{subfigure}{0.18\linewidth}
		\centering
%
 
\begin{tikzpicture}[inner ysep=1.5pt]

\begin{axis}[%
at={(0,0)},
scale only axis,
xlabel style={font=\color{white!15!black}, font=\tiny},
xlabel={density $ L/P $},
xlabel shift = -2.5pt,
xmode = log,
xtick = {0.025,0.05,0.1,0.2,0.4},
xticklabels = {0.025,0.05,0.1,0.2,0.4},
ylabel style={font=\color{white!15!black}, font=\tiny},
ylabel={number of iterations},
ylabel shift = -3pt,
scaled y ticks= base 10:-3,
tick scale binop = \times,
axis background/.style={fill=white},
xmajorgrids,
ymajorgrids,
legend style={legend cell align=left, align=left, draw=white!15!black, 
	font=\tiny, row sep=-2pt, 
	at={(0.5,1)}, anchor = north},
legend columns = 3,
cycle multi list = {
	solid,dashed,densely dotted\nextlist
	[2 of]MATLABcolormarklist
},
]
\addplot table [x=density, y={P=32}] {results/gauss_spatial_no_temporal/mediumScaleTest/varyL/Niter_FUN_PRDLsca.csv};
\addplot table [x=density, y={P=64}] {results/gauss_spatial_no_temporal/mediumScaleTest/varyL/Niter_FUN_PRDLsca.csv};
\addplot table [x=density, y={P=32}] {results/gauss_spatial_no_temporal/mediumScaleTest/varyL/Niter_FUN_PRDLscaX.csv};
\addplot table [x=density, y={P=64}] {results/gauss_spatial_no_temporal/mediumScaleTest/varyL/Niter_FUN_PRDLscaX.csv};
\addplot table [x=density, y={P=32}] {results/gauss_spatial_no_temporal/mediumScaleTest/varyL/Niter_scPRIME.csv};
\addplot table [x=density, y={P=64}] {results/gauss_spatial_no_temporal/mediumScaleTest/varyL/Niter_scPRIME.csv};
\end{axis}
\end{tikzpicture}%
	\end{subfigure}
	\hfill
	\begin{subfigure}{0.18\linewidth}
		\centering
%
 
\begin{tikzpicture}[inner ysep=1.5pt]

\begin{axis}[%
at={(0,0)},
scale only axis,
xlabel style={font=\color{white!15!black}, font=\tiny},
xlabel={density $ L/P $},
xlabel shift = -2.5pt,
xmode = log,
xtick = {0.025,0.05,0.1,0.2,0.4},
xticklabels = {0.025,0.05,0.1,0.2,0.4},
ylabel style={font=\color{white!15!black}, font=\tiny},
ylabel={CPU time (seconds)},
ylabel shift = -3pt,
scaled y ticks= base 10:-2,
tick scale binop = \times,
axis background/.style={fill=white},
xmajorgrids,
ymajorgrids,
legend style={legend cell align=left, align=left, draw=white!15!black, 
	font=\tiny, row sep=-2pt, 
	at={(0.5,1)}, anchor = north},
legend columns = 3,
cycle multi list = {
	solid,dashed,densely dotted\nextlist
	[2 of]MATLABcolormarklist
},
]
\addplot table [x=density, y={P=32}] {results/gauss_spatial_no_temporal/mediumScaleTest/varyL/runtime_FUN_PRDLsca.csv};
\addplot table [x=density, y={P=64}] {results/gauss_spatial_no_temporal/mediumScaleTest/varyL/runtime_FUN_PRDLsca.csv};
\addplot table [x=density, y={P=32}] {results/gauss_spatial_no_temporal/mediumScaleTest/varyL/runtime_FUN_PRDLscaX.csv};
\addplot table [x=density, y={P=64}] {results/gauss_spatial_no_temporal/mediumScaleTest/varyL/runtime_FUN_PRDLscaX.csv};
\addplot table [x=density, y={P=32}] {results/gauss_spatial_no_temporal/mediumScaleTest/varyL/runtime_scPRIME.csv};
\addplot table [x=density, y={P=64}] {results/gauss_spatial_no_temporal/mediumScaleTest/varyL/runtime_scPRIME.csv};
\end{axis}
\end{tikzpicture}%
	\end{subfigure}
	\hfill
	\caption{Performance vs. density $L/P$ using compact-SCAphase (solid), SCAphase (dashed), and SC-PRIME (dotted) in Case~1 with $N \!=\! 64, M_1 \!=\! 4N, I \!=\! 16N$.
	}
	\label{fig:accuracy_varyL}
\end{figure*}

\subsection{Hyperparameter Choices}
\label{subsec:hypeparameter}

\textbf{Sparsity parameter of the cPRDL problem in~\eqref{prob:PRDLMatrixForm}.}
The solution for $\bZ$ in problem~\eqref{prob:PRDLMatrixForm} tends to $\bzero$ as $\lambda \rightarrow \infty$ and there exists an upper bound $\lambda_{\max}$ such that, for $\lambda\! \geq \! \lambda_{\max}$, any point with $\bZ\! = \!\bzero$ is stationary for problem~\eqref{prob:PRDLMatrixForm}~\cite{kimInteriorPointMethodLargeScale2007}. With knowledge of $\lambda_{\max}$, the problem of searching for a suitable sparsity regularization parameter $\lambda$ for an instance is significantly reduced, since any $\lambda \geq \lambda_{\max}$ is ineffective.

From the stationarity conditions~\eqref{eq:stationaryCondition}, an upper bound
\begin{equation}
	\label{eq:lambda_max_general}
	\begin{matrix}
		\lambda_{\max} = \norm{\bY}_\tF \cdot \max_{i=1,\ldots,I} \{ \sigma_{\max}(\bF_i) \}
	\end{matrix}
\end{equation}
can be derived, where $\sigma_{\max}(\cdot)$ denotes the largest singular value.
For $\lambda \! \geq \! \lambda_{\max}$, any point $(\bD,\bzero)$ with $\bD \in \set{D}$ is stationary for the original problem~\eqref{prob:PRDLMatrixForm}. 
Moreover, it is easy to verify that all points $(\bD,\bzero)$ with $\bD \! \in \! \set{D}$ are equally optimal for problem~\eqref{prob:PRDLMatrixForm}.

For the three investigated cases of linear operator $\opr{F}$, $\lambda_{\max}$ can be further decreased.
In Case 1 and 2, where the spatial mixing is time-invariant, $\lambda_{\max}$ can be decreased to
\begin{equation}
	\label{eq:lambda_max_time-invariant}
	\begin{matrix}
		\lambda_{\max} = \sigma_{\max}(\bA) \cdot \max_{i=1,\ldots,I} \big\{ \sum_{m=1}^{M_2} \Abs{b_{i,m}} \cdot \Norm{\by_m}_2 \big\}
	\end{matrix}.
\end{equation}
Then, in Case 3, $\lambda_{\max}$ can be decreased to
\begin{equation}
	\label{eq:lambda_max_withoutTemporalMixing}
	\begin{matrix}
		\lambda_{\max} = \max_{i=1,\ldots,I} \left\{ \sigma_{\max} (\bA_i) \cdot \|\by_i\|_2 \right\}
	\end{matrix}.
\end{equation}

The intermediate derivations of the upper bounds $\lambda_{\max}$ in~\eqref{eq:lambda_max_general}-\eqref{eq:lambda_max_withoutTemporalMixing} can be found in Appendix~\ref{appendix:lambda_max}.

\textbf{Regularization parameters of the PRDL problem in~\eqref{prob:SC-PRIME}.}
Problem~\eqref{prob:SC-PRIME} has two regularization parameters $\mu$ and $\rho$. Similar to $\lambda$ in~\eqref{prob:PRDLMatrixForm}, $\rho$ adjusts the sparsity level of matrix $\bZ$, whereas $\mu$ controls the trade-off between the data fidelity and the approximation quality of the sparse representation.

Similarly, for the sparsity parameter $\rho$ in~\eqref{prob:SC-PRIME}, there exists an upper bound $\rho_{\max}$ such that, for any $\rho \! \geq \! \rho_{\max}$, problem~\eqref{prob:SC-PRIME} always admits a stationary point with $\bZ \!=\! \bzero$.
From the stationarity conditions~\eqref{eq:stationaryCondition_2}-\eqref{eq:stationaryConditionX1}, we obtain an upper bound
\begin{equation}
	\label{eq:rho_max_general}
	\begin{matrix}
		\rho_{\max} = {\mu \cdot \sigma_{\max}(\bF) \cdot \norm{\bY}_\tF} \big/ {\big(\sigma_{\min}^2(\bF)+ \mu\big)}
	\end{matrix}.
\end{equation}
$\sigma_{\min}(\cdot)$ denotes the smallest singular value, which may be zero.
Furthermore, in Case 1 and 3, where no temporal mixing is applied, each snapshot $\bx_i$ is observed independently and, hence, the upper bound $\rho_{\max}$ can be decreased to
\begin{equation}
	\label{eq:rho_max_case3}
		\rho_{\max} = \max_{i=1,\ldots,I} \big\{ {\mu \cdot \sigma_{\max}(\bA_i) \cdot \norm{\by_i}_2} \big/ {\big(\sigma^2_{\min} (\bA_i) + \mu\big)} \big\}.
\end{equation}
Note that Case 1 can be viewed as a special case of Case 3 where $ \bA_i = \bA $ for all snapshots.
The derivations of the upper bounds $\rho_{\max}$ in~\eqref{eq:rho_max_general}-\eqref{eq:rho_max_case3} are provided in Appendix~\ref{appendix:rho_max}.

Next, to analyze the effect of parameter $\mu$, we write the gradient $\nabla_{\! \bX} \widehat{f}$ as
	$ \nabla_{\! \vec (\bX)} \widehat{f} (\bX, \bD, \bZ;\vect{S}^{(t)}) = \big( \bF^\tH \bF + \mu \bI_{NI} \big) \vec (\bX) 
	- \big( \bF^\tH \vec (\bY^{(t)}) + \mu \vec(\bD \bZ) \big) $
with the vectorized form in~\eqref{eq:linearOpr_vectorized}.
Then the stationarity condition~\eqref{eq:stationaryConditionX1} can be rewritten as
\begin{multline}
	\label{eq:stationaryConditionX1_2}
	\vec (\bX) = ( \bF^\tH \bF + \mu \bI_{NI} )^{-1} \bF^\tH \vec (\bY^{(t)}) \\
	+ \big( \tfrac{1}{\mu} \bF^\tH \bF + \bI_{NI} \big)^{-1} \vec(\bD \bZ).
\end{multline}
As shown in~\eqref{eq:stationaryConditionX1_2}, $\mu$ offers some control over how much the value of $ \vec (\bX) $ at a stationary point of $ \widehat{h} $ is influenced by the data fitting solution $ \bF^\dagger \vec (\bY^{(t)}) $ and the sparse representation $ \vec (\bD \bZ) $.
Also, the trade-off depends on both $\mu$ and $ \bF^\tH \bF $.
Thus, we propose to set $\mu$ to be proportional to $ \sigma_{\min,\operatorname{nz}}^2 (\bF) $, where $\sigma_{\min,\operatorname{nz}}(\cdot)$ denotes the smallest nonzero singular value. However, a suitable ratio has to be found by experiments.




If training data are available, one can quickly obtain the suitable values of the regularization parameters by grid search with the upper bound $ \lambda_{\max} $~($\rho_{\max}$) derived above, which is how we choose the regularization parameters in our simulations.
Advanced approaches such as Expectation-Maximization-based methods~\cite{tolooshamsDeepResidualAutoencoders2021} may be applied for simultaneous estimation of hyperparameters, which is subject of future research.

\subsection{Computational Experiments}

In the following, we evaluate the complexity and estimation accuracy of the proposed algorithms under various parameter setups, in comparison to SC-PRIME.
The number of receive antennas is set to $N \!=\! 64$. 
The algorithms are terminated when the minimum-norm subgradient has achieved the tolerance $\varepsilon \!=\! 10^{-5}$ or after a maximum number of 2000 iterations. A following debiasing step is performed with the same termination condition.
By default, the SNR is 15 dB, the spatial over-sampling rate is $M_1/N \!=\! 4$, and $I \!=\! 16N$ time-slots are taken.

\subsubsection{Case 1 -- Time-invariant spatial mixing and no temporal mixing}
\label{subsubsec:case1}

We first consider the case without temporal mixing.
The regularization parameters are set as follows:
$\mu \!=\! \sigma_{\min,\operatorname{nz}}^2(\bF) \!=\! \sigma_{\min,\operatorname{nz}}^2 (\bA)$ for both SCAphase and SC-PRIME,
$\lambda \!=\! 0.75^{16} \lambda_{\max}$ with $\lambda_{\max}$ in~\eqref{eq:lambda_max_time-invariant} for compact-SCAphase, and $ \rho \!=\! 0.75^{16} \rho_{\max} $ with $\rho_{\max}$ in~\eqref{eq:rho_max_case3} for SCAphase. Although SC-PRIME adopts the same formulation, i.e., problem~\eqref{prob:SC-PRIME}, as SCAphase, it typically requires a larger sparsity parameter $ \rho $ for achieving a good solution, due to the loose majorization on the data fitting term employed in the surrogate subproblems. Thus, for SC-PRIME, $\rho$ is set to be $ 0.75^{15} \rho_{\max} $ and $ 0.75^{14} \rho_{\max}$ in the cases with $ P \!=\! N/2 $ and $ P \!=\! N $, respectively.

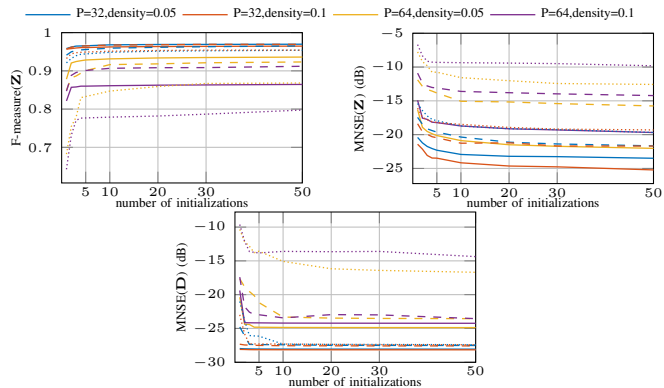
\begin{figure}[t]
	\centering
	\ref{legend3}
	\vspace*{-3pt}
	
	\begin{subfigure}{0.48\linewidth}
		\centering
%
\begin{tikzpicture}[inner ysep=1.5pt]

\begin{axis}[%
width=3.2cm,
height=2cm,
at={(0,0)},
scale only axis,
xmin=0,
xmax=50,
xlabel style={font=\color{white!15!black}, font=\tiny},
xlabel={number of initializations},
xlabel shift = -2.5pt,
xtick = {5,10,20,30,50},
ymax=1,
ylabel style={font=\color{white!15!black}, font=\tiny},
ylabel={F-measure($\bZ$)},
ylabel shift = -3pt,
axis background/.style={fill=white},
xmajorgrids,
ymajorgrids,
legend style={legend cell align=left, align=left, draw=white!15!black, 
	font=\tiny, row sep=-2pt, draw=none,
	at={(1,0)}, anchor = south east},
legend columns = 4,
cycle multi list = {
	solid,dashed,densely dotted\nextlist
	[4 of]MATLABcolorlist
},
legend to name=legend3,
]

\addplot table [col sep=tab, x={no. of initializations}, y={P=32,density=0.05}] {results/gauss_spatial_no_temporal/mediumScaleTest/varyInitial/FmeasureZ_FUN_PRDLsca.csv};
\addplot table [col sep=tab, x={no. of initializations}, y={P=32,density=0.10}] {results/gauss_spatial_no_temporal/mediumScaleTest/varyInitial/FmeasureZ_FUN_PRDLsca.csv};
\addplot table [col sep=tab, x={no. of initializations}, y={P=64,density=0.05}] {results/gauss_spatial_no_temporal/mediumScaleTest/varyInitial/FmeasureZ_FUN_PRDLsca.csv};
\addplot table [col sep=tab, x={no. of initializations}, y={P=64,density=0.10}] {results/gauss_spatial_no_temporal/mediumScaleTest/varyInitial/FmeasureZ_FUN_PRDLsca.csv};

\addplot table [col sep=tab, x={no. of initializations}, y={P=32,density=0.05}] {results/gauss_spatial_no_temporal/mediumScaleTest/varyInitial/FmeasureZ_FUN_PRDLscaX.csv};
\addplot table [col sep=tab, x={no. of initializations}, y={P=32,density=0.10}] {results/gauss_spatial_no_temporal/mediumScaleTest/varyInitial/FmeasureZ_FUN_PRDLscaX.csv};
\addplot table [col sep=tab, x={no. of initializations}, y={P=64,density=0.05}] {results/gauss_spatial_no_temporal/mediumScaleTest/varyInitial/FmeasureZ_FUN_PRDLscaX.csv};
\addplot table [col sep=tab, x={no. of initializations}, y={P=64,density=0.10}] {results/gauss_spatial_no_temporal/mediumScaleTest/varyInitial/FmeasureZ_FUN_PRDLscaX.csv};

\addplot table [col sep=tab, x={no. of initializations}, y={P=32,density=0.05}] {results/gauss_spatial_no_temporal/mediumScaleTest/varyInitial/FmeasureZ_scPRIME.csv};
\addplot table [col sep=tab, x={no. of initializations}, y={P=32,density=0.10}] {results/gauss_spatial_no_temporal/mediumScaleTest/varyInitial/FmeasureZ_scPRIME.csv};
\addplot table [col sep=tab, x={no. of initializations}, y={P=64,density=0.05}] {results/gauss_spatial_no_temporal/mediumScaleTest/varyInitial/FmeasureZ_scPRIME.csv};
\addplot table [col sep=tab, x={no. of initializations}, y={P=64,density=0.10}] {results/gauss_spatial_no_temporal/mediumScaleTest/varyInitial/FmeasureZ_scPRIME.csv};

\legend{{P=32,density=0.05},{P=32,density=0.1},{P=64,density=0.05},{P=64,density=0.1}};
\end{axis}
\end{tikzpicture}%
	\end{subfigure}
	\hfill
	\begin{subfigure}{0.48\linewidth}
		\centering
%
\begin{tikzpicture}[inner ysep=1.5pt]

\begin{axis}[%
width=3.2cm,
height=2cm,
at={(0,0)},
scale only axis,
xmin=0,
xmax=50,
xlabel style={font=\color{white!15!black}, font=\tiny},
xlabel={number of initializations},
xlabel shift = -2.5pt,
xtick = {5,10,20,30,50},
ymax=-5,
ylabel style={font=\color{white!15!black}, font=\tiny},
ylabel={MNSE($\bZ$) (dB)},
ylabel shift = -3pt,
axis background/.style={fill=white},
xmajorgrids,
ymajorgrids,
legend style={legend cell align=left, align=left, draw=white!15!black, 
	font=\tiny, row sep=-2pt, draw=none,
	at={(1,0)}, anchor = south east},
legend columns = 4,
cycle multi list = {
	solid,dashed,densely dotted\nextlist
	[4 of]MATLABcolorlist
},
legend to name=legend3,
]

\addplot table [col sep=tab, x={no. of initializations}, y={P=32,density=0.05}] {results/gauss_spatial_no_temporal/mediumScaleTest/varyInitial/errorZ_FUN_PRDLsca.csv};
\addplot table [col sep=tab, x={no. of initializations}, y={P=32,density=0.10}] {results/gauss_spatial_no_temporal/mediumScaleTest/varyInitial/errorZ_FUN_PRDLsca.csv};
\addplot table [col sep=tab, x={no. of initializations}, y={P=64,density=0.05}] {results/gauss_spatial_no_temporal/mediumScaleTest/varyInitial/errorZ_FUN_PRDLsca.csv};
\addplot table [col sep=tab, x={no. of initializations}, y={P=64,density=0.10}] {results/gauss_spatial_no_temporal/mediumScaleTest/varyInitial/errorZ_FUN_PRDLsca.csv};

\addplot table [col sep=tab, x={no. of initializations}, y={P=32,density=0.05}] {results/gauss_spatial_no_temporal/mediumScaleTest/varyInitial/errorZ_FUN_PRDLscaX.csv};
\addplot table [col sep=tab, x={no. of initializations}, y={P=32,density=0.10}] {results/gauss_spatial_no_temporal/mediumScaleTest/varyInitial/errorZ_FUN_PRDLscaX.csv};
\addplot table [col sep=tab, x={no. of initializations}, y={P=64,density=0.05}] {results/gauss_spatial_no_temporal/mediumScaleTest/varyInitial/errorZ_FUN_PRDLscaX.csv};
\addplot table [col sep=tab, x={no. of initializations}, y={P=64,density=0.10}] {results/gauss_spatial_no_temporal/mediumScaleTest/varyInitial/errorZ_FUN_PRDLscaX.csv};

\addplot table [col sep=tab, x={no. of initializations}, y={P=32,density=0.05}] {results/gauss_spatial_no_temporal/mediumScaleTest/varyInitial/errorZ_scPRIME.csv};
\addplot table [col sep=tab, x={no. of initializations}, y={P=32,density=0.10}] {results/gauss_spatial_no_temporal/mediumScaleTest/varyInitial/errorZ_scPRIME.csv};
\addplot table [col sep=tab, x={no. of initializations}, y={P=64,density=0.05}] {results/gauss_spatial_no_temporal/mediumScaleTest/varyInitial/errorZ_scPRIME.csv};
\addplot table [col sep=tab, x={no. of initializations}, y={P=64,density=0.10}] {results/gauss_spatial_no_temporal/mediumScaleTest/varyInitial/errorZ_scPRIME.csv};

\legend{{P=32,density=0.05},{P=32,density=0.1},{P=64,density=0.05},{P=64,density=0.1}};
\end{axis}
\end{tikzpicture}%
	\end{subfigure} \\
	\begin{subfigure}{0.48\linewidth}
		\centering
%
\begin{tikzpicture}[inner ysep=1.5pt]

\begin{axis}[%
width=3.2cm,
height=2cm,
at={(0,0)},
scale only axis,
xmin=0,
xmax=50,
xlabel style={font=\color{white!15!black}, font=\tiny},
xlabel={number of initializations},
xlabel shift = -2.5pt,
xtick = {5,10,20,30,50},
ylabel style={font=\color{white!15!black}, font=\tiny},
ylabel={MNSE($\bD$) (dB)},
ylabel shift = -3pt,
axis background/.style={fill=white},
xmajorgrids,
ymajorgrids,
legend style={legend cell align=left, align=left, draw=white!15!black, 
	font=\tiny, row sep=-2pt, draw=none,
	at={(1,0)}, anchor = south east},
legend columns = 4,
cycle multi list = {
	solid,dashed,densely dotted\nextlist
	[4 of]MATLABcolorlist
},
legend to name=legend3,
]

\addplot table [col sep=tab, x={no. of initializations}, y={P=32,density=0.05}] {results/gauss_spatial_no_temporal/mediumScaleTest/varyInitial/errorD_FUN_PRDLsca.csv};
\addplot table [col sep=tab, x={no. of initializations}, y={P=32,density=0.10}] {results/gauss_spatial_no_temporal/mediumScaleTest/varyInitial/errorD_FUN_PRDLsca.csv};
\addplot table [col sep=tab, x={no. of initializations}, y={P=64,density=0.05}] {results/gauss_spatial_no_temporal/mediumScaleTest/varyInitial/errorD_FUN_PRDLsca.csv};
\addplot table [col sep=tab, x={no. of initializations}, y={P=64,density=0.10}] {results/gauss_spatial_no_temporal/mediumScaleTest/varyInitial/errorD_FUN_PRDLsca.csv};

\addplot table [col sep=tab, x={no. of initializations}, y={P=32,density=0.05}] {results/gauss_spatial_no_temporal/mediumScaleTest/varyInitial/errorD_FUN_PRDLscaX.csv};
\addplot table [col sep=tab, x={no. of initializations}, y={P=32,density=0.10}] {results/gauss_spatial_no_temporal/mediumScaleTest/varyInitial/errorD_FUN_PRDLscaX.csv};
\addplot table [col sep=tab, x={no. of initializations}, y={P=64,density=0.05}] {results/gauss_spatial_no_temporal/mediumScaleTest/varyInitial/errorD_FUN_PRDLscaX.csv};
\addplot table [col sep=tab, x={no. of initializations}, y={P=64,density=0.10}] {results/gauss_spatial_no_temporal/mediumScaleTest/varyInitial/errorD_FUN_PRDLscaX.csv};

\addplot table [col sep=tab, x={no. of initializations}, y={P=32,density=0.05}] {results/gauss_spatial_no_temporal/mediumScaleTest/varyInitial/errorD_scPRIME.csv};
\addplot table [col sep=tab, x={no. of initializations}, y={P=32,density=0.10}] {results/gauss_spatial_no_temporal/mediumScaleTest/varyInitial/errorD_scPRIME.csv};
\addplot table [col sep=tab, x={no. of initializations}, y={P=64,density=0.05}] {results/gauss_spatial_no_temporal/mediumScaleTest/varyInitial/errorD_scPRIME.csv};
\addplot table [col sep=tab, x={no. of initializations}, y={P=64,density=0.10}] {results/gauss_spatial_no_temporal/mediumScaleTest/varyInitial/errorD_scPRIME.csv};

\legend{{P=32,density=0.05},{P=32,density=0.1},{P=64,density=0.05},{P=64,density=0.1}};
\end{axis}
\end{tikzpicture}%
	\end{subfigure}
	\hfill
	\caption{Estimation quality vs. number of initializations using compact-SCAphase (solid), SCAphase (dashed), and SC-PRIME (dotted) in Case 1 with $N \!=\! 64, \ M_1 \!=\! 4N, \ I \!=\! 16N$.
	}
	\label{fig:accuracy_varyInitial}
\end{figure}

\textbf{Varying sparsity level.} In the first simulation, as depicted in Fig.~\ref{fig:accuracy_varyL}, the performance of the algorithms is evaluated for various choices of $\{P,L/P\}$. 
The number of users $P$ is varied in $\{N/2,N\}$, and the density of active users in each time-slot, i.e., $L/P$, is limited to be $\{0.025,0.05,0.1,0.2,0.4\}$.
As both problems~\eqref{prob:PRDLMatrixForm} and~\eqref{prob:SC-PRIME} are nonconvex, multiple random initializations are used to increase the chance of finding the global optimal solution. Specifically, for each Monte-Carlo trial, $ 10 $ initializations are performed, and the best reconstructed signal, determined by the lowest objective function value, is retained and further improved by a debiasing step.
The total number of iterations and computational time, including that of the debiasing step, are reported in Fig.~\ref{fig:accuracy_varyL}.
The robustness of the algorithms to initialization is investigated afterwards in Fig.~\ref{fig:accuracy_varyInitial}.

From Fig.~\ref{fig:accuracy_varyL}, it can be observed that sparse channel access is required, i.e., a small value of $L/P$, for all algorithms to achieve good recovery performance. 
However, in the extremely sparse case, the received signals $\bX^\text{true} \!=\! \bD^\text{true} \bZ^\text{true} $ contain only few linear combinations of columns of spatial signature $\bD^\text{true}$, which results in a degradation of estimation qualities.
Furthermore, for all choices of $ \{P,L/P\} $, SC-PRIME does not converge within 2000 iterations. The solution obtained by SC-PRIME within 2000 iterations can be improved by using a larger sparsity parameter $ \rho $ than that in SCAphase, as in the parameter setup of this simulation. However, in Fig.~\ref{fig:accuracy_varyL}, SC-PRIME still exhibits the poorest accuracy performance for most choices of $ \{P,L/P\} $, compared to the other algorithms.

In Fig.~\ref{fig:accuracy_varyL}, when $ P = N/2 $, all algorithms show good recovery performance, whereas compact-SCAphase and SCAphase exhibit faster convergence. Moreover, compared to SCAphase, compact-SCAphase uses half the number of iterations to attain a stationary point.
However, the reduction of CPU time achieved by compact-SCAphase is not as significant as the reduced number of iterations because, as discussed in Section~\ref{subsec:complexity}, compact-SCAphase has the highest per-iteration complexity.
In contrast, when the number of users is comparable to that of antennas, i.e., $P \! = \! N$, only compact-SCAphase achieves the given tolerance within 2000 iterations. This is intuitive as in the regime of $P \! \geq \! N$, and with sparse channel access, the information of the users' channels contained in the measurements is insufficient. To resolve this challenge, a higher spatial oversampling rate is required.
Nevertheless, compared to SC-PRIME, compact-SCAphase and SCAphase show a significant improvement of estimation accuracy. Then, compared to SCAphase, compact-SCAphase further improves the estimation quality of $ \bZ $ due to fast convergence.

\textbf{Varying number of initializations.} 
In the second simulation, we investigate the robustness of the algorithms to initialization. The performance behavior of the algorithms with the number of random initializations varied from $1$ to $50$ is presented in Fig.~\ref{fig:accuracy_varyInitial}.
The number of users $P$ and density are set to be $\{N/2, N\}$ and $\{0.05,0.1\}$, respectively. 
In Fig.~\ref{fig:accuracy_varyInitial}, for most choices of $\{P,L/P\}$, all algorithms show similar robustness to initialization as the estimation quality achieved by each algorithm remains constant after the trial of $ 10 $ initializations,
and compact-SCAphase possesses the lowest estimation errors. 
In the cases with $ P=N $, SC-PRIME shows a significant degradation on the estimation quality compared to the proposed algorithms, which, as demonstrated in Fig.~\ref{fig:accuracy_varyL}, results from the fact that SC-PRIME generally does not converge within the limit of 2000 iterations.
Additionally, if only the spatial signature $ \bD $ needs to be recovered, then $ 5 $ initializations are sufficient for all algorithms to attain a good estimation accuracy. Particularly, when $ P = N/2 $, compact-SCAphase achieves a good stationary point for $ \bD $ even with a single initialization.


\subsubsection{Case 2 -- Time-invariant spatial mixing and STFT temporal mixing}

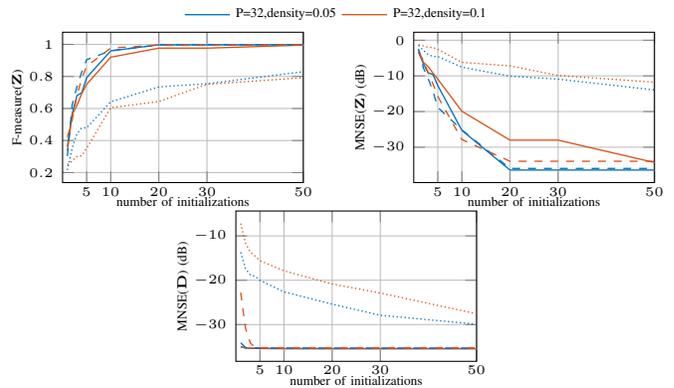
\begin{figure}[t]
	\centering
	\ref{legend4}

	\begin{subfigure}{0.48\linewidth}
		\centering
%
\begin{tikzpicture}[inner ysep=1.5pt]

\begin{axis}[%
width=3.2cm,
height=2cm,
at={(0,0)},
scale only axis,
xmin=0,
xmax=50,
xlabel style={font=\color{white!15!black}, font=\tiny},
xlabel={number of initializations},
xlabel shift = -2.5pt,
xtick = {5,10,20,30,50},
ylabel style={font=\color{white!15!black}, font=\tiny},
ylabel={F-measure($\bZ$)},
ylabel shift = -3pt,
axis background/.style={fill=white},
xmajorgrids,
ymajorgrids,
legend style={legend cell align=left, align=left, draw=white!15!black, 
	font=\tiny, row sep=-2pt, draw=none,
	at={(1,0)}, anchor = south east},
legend columns = 4,
cycle multi list = {
	solid,dashed,densely dotted\nextlist
	[2 of]MATLABcolorlist
},
legend to name=legend4,
]

\addplot table [col sep=tab, x={no. of initializations}, y={P=32,density=0.05}] {results/gauss_spatial_stft_temporal/mediumScaleTest/varyInitial/FmeasureZ_FUN_PRDLsca.csv};
\addplot table [col sep=tab, x={no. of initializations}, y={P=32,density=0.10}] {results/gauss_spatial_stft_temporal/mediumScaleTest/varyInitial/FmeasureZ_FUN_PRDLsca.csv};

\addplot table [col sep=tab, x={no. of initializations}, y={P=32,density=0.05}] {results/gauss_spatial_stft_temporal/mediumScaleTest/varyInitial/FmeasureZ_FUN_PRDLscaX.csv};
\addplot table [col sep=tab, x={no. of initializations}, y={P=32,density=0.10}] {results/gauss_spatial_stft_temporal/mediumScaleTest/varyInitial/FmeasureZ_FUN_PRDLscaX.csv};

\addplot table [col sep=tab, x={no. of initializations}, y={P=32,density=0.05}] {results/gauss_spatial_stft_temporal/mediumScaleTest/varyInitial/FmeasureZ_scPRIME.csv};
\addplot table [col sep=tab, x={no. of initializations}, y={P=32,density=0.10}] {results/gauss_spatial_stft_temporal/mediumScaleTest/varyInitial/FmeasureZ_scPRIME.csv};

\legend{{P=32,density=0.05},{P=32,density=0.1}};
\end{axis}
\end{tikzpicture}%
	\end{subfigure}
	\hfill
	\begin{subfigure}{0.48\linewidth}
		\centering
%
\begin{tikzpicture}[inner ysep=1.5pt]

\begin{axis}[%
width=3.2cm,
height=2cm,
at={(0,0)},
scale only axis,
xmin=0,
xmax=50,
xlabel style={font=\color{white!15!black}, font=\tiny},
xlabel={number of initializations},
xlabel shift = -2.5pt,
xtick = {5,10,20,30,50},
ylabel style={font=\color{white!15!black}, font=\tiny},
ylabel={MNSE($\bZ$) (dB)},
ylabel shift = -3pt,
axis background/.style={fill=white},
xmajorgrids,
ymajorgrids,
legend style={legend cell align=left, align=left, draw=white!15!black, 
	font=\tiny, row sep=-2pt, draw=none,
	at={(1,0)}, anchor = south east},
legend columns = 4,
cycle multi list = {
	solid,dashed,densely dotted\nextlist
	[2 of]MATLABcolorlist
},
legend to name=legend4,
]

\addplot table [col sep=tab, x={no. of initializations}, y={P=32,density=0.05}] {results/gauss_spatial_stft_temporal/mediumScaleTest/varyInitial/errorZ_FUN_PRDLsca.csv};
\addplot table [col sep=tab, x={no. of initializations}, y={P=32,density=0.10}] {results/gauss_spatial_stft_temporal/mediumScaleTest/varyInitial/errorZ_FUN_PRDLsca.csv};

\addplot table [col sep=tab, x={no. of initializations}, y={P=32,density=0.05}] {results/gauss_spatial_stft_temporal/mediumScaleTest/varyInitial/errorZ_FUN_PRDLscaX.csv};
\addplot table [col sep=tab, x={no. of initializations}, y={P=32,density=0.10}] {results/gauss_spatial_stft_temporal/mediumScaleTest/varyInitial/errorZ_FUN_PRDLscaX.csv};

\addplot table [col sep=tab, x={no. of initializations}, y={P=32,density=0.05}] {results/gauss_spatial_stft_temporal/mediumScaleTest/varyInitial/errorZ_scPRIME.csv};
\addplot table [col sep=tab, x={no. of initializations}, y={P=32,density=0.10}] {results/gauss_spatial_stft_temporal/mediumScaleTest/varyInitial/errorZ_scPRIME.csv};

\legend{{P=32,density=0.05},{P=32,density=0.1}};
\end{axis}
\end{tikzpicture}%
	\end{subfigure}\\
	\begin{subfigure}{0.48\linewidth}
		\centering
%
\begin{tikzpicture}[inner ysep=1.5pt]

\begin{axis}[%
width=3.2cm,
height=2cm,
at={(0,0)},
scale only axis,
xmin=0,
xmax=50,
xlabel style={font=\color{white!15!black}, font=\tiny},
xlabel={number of initializations},
xlabel shift = -2.5pt,
xtick = {5,10,20,30,50},
ylabel style={font=\color{white!15!black}, font=\tiny},
ylabel={MNSE($\bD$) (dB)},
ylabel shift = -3pt,
axis background/.style={fill=white},
xmajorgrids,
ymajorgrids,
legend style={legend cell align=left, align=left, draw=white!15!black, 
	font=\tiny, row sep=-2pt, draw=none,
	at={(1,0)}, anchor = south east},
legend columns = 4,
cycle multi list = {
	solid,dashed,densely dotted\nextlist
	[2 of]MATLABcolorlist
},
legend to name=legend4,
]

\addplot table [col sep=tab, x={no. of initializations}, y={P=32,density=0.05}] {results/gauss_spatial_stft_temporal/mediumScaleTest/varyInitial/errorD_FUN_PRDLsca.csv};
\addplot table [col sep=tab, x={no. of initializations}, y={P=32,density=0.10}] {results/gauss_spatial_stft_temporal/mediumScaleTest/varyInitial/errorD_FUN_PRDLsca.csv};

\addplot table [col sep=tab, x={no. of initializations}, y={P=32,density=0.05}] {results/gauss_spatial_stft_temporal/mediumScaleTest/varyInitial/errorD_FUN_PRDLscaX.csv};
\addplot table [col sep=tab, x={no. of initializations}, y={P=32,density=0.10}] {results/gauss_spatial_stft_temporal/mediumScaleTest/varyInitial/errorD_FUN_PRDLscaX.csv};

\addplot table [col sep=tab, x={no. of initializations}, y={P=32,density=0.05}] {results/gauss_spatial_stft_temporal/mediumScaleTest/varyInitial/errorD_scPRIME.csv};
\addplot table [col sep=tab, x={no. of initializations}, y={P=32,density=0.10}] {results/gauss_spatial_stft_temporal/mediumScaleTest/varyInitial/errorD_scPRIME.csv};

\legend{{P=32,density=0.05},{P=32,density=0.1}};
\end{axis}
\end{tikzpicture}%
	\end{subfigure}
	\hfill
	\caption{Estimation quality vs. number of initializations using compact-SCAphase (solid), SCAphase (dashed), and SC-PRIME (dotted) in Case 2 with $N \!=\! 64, \ M_1 \!=\! 4N, \ I \!=\! 16N$.
	}
	\label{fig:accuracy_withTemporalMixing}
\end{figure}

Next, a temporal mixing network that performs the same STFT independently on each output channel of the spatial mixing network is introduced (see~\cite{jaganathanSTFTPhaseRetrieval2016} for more details of the STFT measurement model). For the STFT, we use an $I$-point DFT, a rectangular window of length $I/2$, a hop size of $I/4$. The above parameter setup results in a temporal oversampling rate of $5$.
Similar to the previous simulation, in Fig.~\ref{fig:accuracy_withTemporalMixing}, the estimation accuracy of the algorithms is evaluated as a function of number of initializations.
We set $P \!=\! N/2$ and density $L/P \!=\! \{0.05,0.1\}$. The regularization parameters are chosen to be $\lambda \!=\! 0.75^{25} \lambda_{\max},\ \mu \!=\! \sigma_{\min,\operatorname{nz}}^2(\bF) \!=\! \sigma_{\min,\operatorname{nz}}^2 (\bA) \sigma_{\min,\operatorname{nz}}^2 (\bB)$, and $\rho \!=\! 0.75^{28} \rho_{\max}$ and $ \rho \! = \! 0.75^{23} \rho_{\max} $ for SCAphase and SC-PRIME, respectively, with $\rho_{\max}$ in~\eqref{eq:rho_max_general}.

Comparing the results in Fig.~\ref{fig:accuracy_varyInitial} and~\ref{fig:accuracy_withTemporalMixing}, we observe that, given a sufficient number of initializations, the estimation qualities are significantly improved in the case with STFT temporal mixing due to the increase of overall sampling rate. 
However, all algorithms become less robust to initialization. In particular, compact-SCAphase and SCAphase require $20$ initializations to attain a good stationary point, whereas SC-PRIME cannot achieve the same estimation accuracy as the other algorithms even with $ 50 $ initializations since, as we discussed, SC-PRIME does not converge within 2000 iterations.

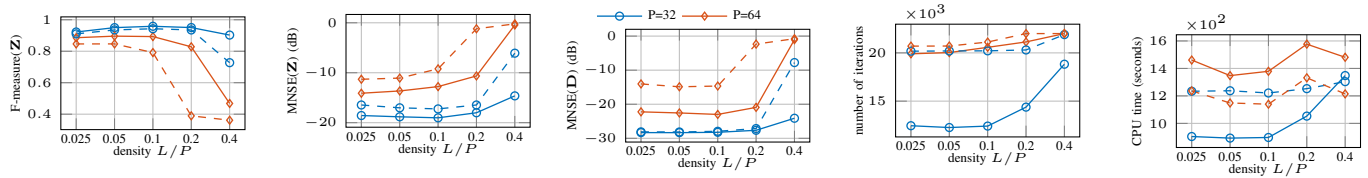
\begin{figure*}[t]
	\centering
	
	\begin{subfigure}{0.18\linewidth}
		\centering
%
%

\begin{tikzpicture}[inner ysep=1.5pt]

\begin{axis}[%
at={(0,0)},
scale only axis,
xlabel style={font=\color{white!15!black}, font=\tiny},
xlabel={density $ L/P $},
xlabel shift = -2.5pt,
xmode = log,
xtick = {0.025,0.05,0.1,0.2,0.4},
xticklabels = {0.025,0.05,0.1,0.2,0.4},
ylabel style={font=\color{white!15!black}, font=\tiny},
ylabel={F-measure($\bZ$)},
ylabel shift = -3pt,
axis background/.style={fill=white},
xmajorgrids,
ymajorgrids,
legend style={legend cell align=left, align=left, draw=white!15!black, 
	font=\tiny, row sep=-2pt, draw=none,
	at={(0,0)}, anchor = south west},
cycle multi list = {
	solid,dashed,densely dotted\nextlist
	[2 of]MATLABcolormarklist
},
]

\addplot table [x=density, y={P=32}] {results/time_variant_gauss_spatial_no_temporal/mediumScaleTest/varyL/FmeasureZ_FUN_PRDLscaX.csv};
\addplot table [x=density, y={P=64}] {results/time_variant_gauss_spatial_no_temporal/mediumScaleTest/varyL/FmeasureZ_FUN_PRDLscaX.csv};
\addplot table [x=density, y={P=32}] {results/time_variant_gauss_spatial_no_temporal/mediumScaleTest/varyL/FmeasureZ_scPRIME.csv};
\addplot table [x=density, y={P=64}] {results/time_variant_gauss_spatial_no_temporal/mediumScaleTest/varyL/FmeasureZ_scPRIME.csv};

\end{axis}
\end{tikzpicture}%
	\end{subfigure}
	\hfill
	\begin{subfigure}{0.18\linewidth}
		\centering
%
%

\begin{tikzpicture}[inner ysep=1.5pt]

\begin{axis}[%
at={(0,0)},
scale only axis,
xlabel style={font=\color{white!15!black}, font=\tiny},
xlabel={density $ L/P $},
xlabel shift = -2.5pt,
xmode = log,
xtick = {0.025,0.05,0.1,0.2,0.4},
xticklabels = {0.025,0.05,0.1,0.2,0.4},
ylabel style={font=\color{white!15!black}, font=\tiny},
ylabel={MNSE($\bZ$) (dB)},
ylabel shift = -3pt,
axis background/.style={fill=white},
xmajorgrids,
ymajorgrids,
legend style={legend cell align=left, align=left, draw=white!15!black, 
	font=\tiny, row sep=-2pt, draw=none,
	at={(0,0)}, anchor = south west},
cycle multi list = {
	solid,dashed,densely dotted\nextlist
	[2 of]MATLABcolormarklist
},
]

\addplot table [x=density, y={P=32}] {results/time_variant_gauss_spatial_no_temporal/mediumScaleTest/varyL/errorZ_FUN_PRDLscaX.csv};
\addplot table [x=density, y={P=64}] {results/time_variant_gauss_spatial_no_temporal/mediumScaleTest/varyL/errorZ_FUN_PRDLscaX.csv};
\addplot table [x=density, y={P=32}] {results/time_variant_gauss_spatial_no_temporal/mediumScaleTest/varyL/errorZ_scPRIME.csv};
\addplot table [x=density, y={P=64}] {results/time_variant_gauss_spatial_no_temporal/mediumScaleTest/varyL/errorZ_scPRIME.csv};

\end{axis}
\end{tikzpicture}%
	\end{subfigure}
	\hfill
	\begin{subfigure}{0.18\linewidth}
		\centering
		\ref{legend2}
		\vspace*{-2pt}
%
%

\begin{tikzpicture}[inner ysep=1.5pt]

\begin{axis}[%
at={(0,0)},
scale only axis,
xlabel style={font=\color{white!15!black}, font=\tiny},
xlabel={density $ L/P $},
xlabel shift = -2.5pt,
xmode = log,
xtick = {0.025,0.05,0.1,0.2,0.4},
xticklabels = {0.025,0.05,0.1,0.2,0.4},
ylabel style={font=\color{white!15!black}, font=\tiny},
ylabel={MNSE($\bD$) (dB)},
ylabel shift = -3pt,
axis background/.style={fill=white},
xmajorgrids,
ymajorgrids,
legend style={legend cell align=left, align=left, draw=white!15!black, 
	font=\tiny, row sep=-2pt, draw=none,
	at={(0,0)}, anchor = south west},
cycle multi list = {
	solid,dashed,densely dotted\nextlist
	[2 of]MATLABcolormarklist
},
]

\addplot table [x=density, y={P=32}] {results/time_variant_gauss_spatial_no_temporal/mediumScaleTest/varyL/errorD_FUN_PRDLscaX.csv};
\addplot table [x=density, y={P=64}] {results/time_variant_gauss_spatial_no_temporal/mediumScaleTest/varyL/errorD_FUN_PRDLscaX.csv};
\addplot table [x=density, y={P=32}] {results/time_variant_gauss_spatial_no_temporal/mediumScaleTest/varyL/errorD_scPRIME.csv};
\addplot table [x=density, y={P=64}] {results/time_variant_gauss_spatial_no_temporal/mediumScaleTest/varyL/errorD_scPRIME.csv};

\end{axis}
\end{tikzpicture}%
	\end{subfigure}
	\hfill
	\begin{subfigure}{0.18\linewidth}
		\centering
%
 
\begin{tikzpicture}[inner ysep=1.5pt]

\begin{axis}[%
at={(0,0)},
scale only axis,
xlabel style={font=\color{white!15!black}, font=\tiny},
xlabel={density $ L/P $},
xlabel shift = -2.5pt,
xmode = log,
xtick = {0.025,0.05,0.1,0.2,0.4},
xticklabels = {0.025,0.05,0.1,0.2,0.4},
ylabel style={font=\color{white!15!black}, font=\tiny},
ylabel={number of iterations},
ylabel shift = -3pt,
scaled y ticks= base 10:-3,
tick scale binop = \times,
axis background/.style={fill=white},
xmajorgrids,
ymajorgrids,
legend style={legend cell align=left, align=left, draw=white!15!black, 
	font=\tiny, row sep=-2pt, 
	at={(0.5,1)}, anchor = north},
legend columns = 3,
cycle multi list = {
	solid,dashed,densely dotted\nextlist
	[2 of]MATLABcolormarklist
},
]

\addplot table [x=density, y={P=32}] {results/time_variant_gauss_spatial_no_temporal/mediumScaleTest/varyL/Niter_FUN_PRDLscaX.csv};
\addplot table [x=density, y={P=64}] {results/time_variant_gauss_spatial_no_temporal/mediumScaleTest/varyL/Niter_FUN_PRDLscaX.csv};
\addplot table [x=density, y={P=32}] {results/time_variant_gauss_spatial_no_temporal/mediumScaleTest/varyL/Niter_scPRIME.csv};
\addplot table [x=density, y={P=64}] {results/time_variant_gauss_spatial_no_temporal/mediumScaleTest/varyL/Niter_scPRIME.csv};
\end{axis}
\end{tikzpicture}%
	\end{subfigure}
	\hfill
	\begin{subfigure}{0.18\linewidth}
		\centering
		\vspace{8pt}
		
%
 
\begin{tikzpicture}[inner ysep=1.5pt]

\begin{axis}[%
at={(0,0)},
scale only axis,
xlabel style={font=\color{white!15!black}, font=\tiny},
xlabel={density $ L/P $},
xlabel shift = -2.5pt,
xmode = log,
xtick = {0.025,0.05,0.1,0.2,0.4},
xticklabels = {0.025,0.05,0.1,0.2,0.4},
ylabel style={font=\color{white!15!black}, font=\tiny},
ylabel={CPU time (seconds)},
ylabel shift = -3pt,
scaled y ticks= base 10:-2,
tick scale binop = \times,
axis background/.style={fill=white},
xmajorgrids,
ymajorgrids,
legend style={legend cell align=left, align=left, draw=white!15!black, 
	font=\tiny, row sep=-2pt, 
	at={(0.5,1)}, anchor = north},
legend columns = 3,
cycle multi list = {
	solid,dashed,densely dotted\nextlist
	[2 of]MATLABcolormarklist
},
]

\addplot table [x=density, y={P=32}] {results/time_variant_gauss_spatial_no_temporal/mediumScaleTest/varyL/runtime_FUN_PRDLscaX.csv};
\addplot table [x=density, y={P=64}] {results/time_variant_gauss_spatial_no_temporal/mediumScaleTest/varyL/runtime_FUN_PRDLscaX.csv};
\addplot table [x=density, y={P=32}] {results/time_variant_gauss_spatial_no_temporal/mediumScaleTest/varyL/runtime_scPRIME.csv};
\addplot table [x=density, y={P=64}] {results/time_variant_gauss_spatial_no_temporal/mediumScaleTest/varyL/runtime_scPRIME.csv};
\end{axis}
\end{tikzpicture}%
	\end{subfigure}
	\caption{Performance vs. density $L/P$ using SCAphase (solid) and SC-PRIME (dashed) in Case 3 with $N \!=\! 64, \ M_1 \!=\! 4N, \ I \!=\! 16N$.
	}
	\label{fig:accuracy_varyL_time-variant}
\end{figure*}

\subsubsection{Case 3 -- Time-variant spatial mixing and no temporal mixing}

As discussed in Section~\ref{subsec:complexity}, compared to the other two algorithms, compact-SCAphase has a per-iteration complexity of higher order in the general case with a linear measurement operator $ \opr{F} $ in~\eqref{eq:oprF_general} with multiple chains of mixing networks.
Therefore, in the case with time-variant spatial mixing, we only compare SCAphase with SC-PRIME, as the running time of compact-SCAphase is unaffordable.
As depicted in Fig.~\ref{fig:accuracy_varyL_time-variant}, the accuracy and complexity of the algorithms are evaluated for various choices of $\{P,L/P\}$. All parameters are the same as in Fig.~\ref{fig:accuracy_varyL}, except that the spatial mixing $\bA_i$ for each snapshot is generated independently. 

What stands out in Fig.~\ref{fig:accuracy_varyL_time-variant} is that the use of time-variant spatial mixing overcomes the challenge of lack of diversity in the extremely sparse case observed in Fig.~\ref{fig:accuracy_varyL}. 
On the other hand, the convergence rates of the two algorithms measured by number of iterations in Fig.~\ref{fig:accuracy_varyL_time-variant} are similar to that in Fig.~\ref{fig:accuracy_varyL}. However, due to the increased complexity of linear operator $ \opr{F} $, the two algorithms possess similar per-iteration complexity. Hence, compared to SC-PRIME, SCAphase exhibits a significantly improved convergence rate in terms of both number of iterations and computational time, when $ P \!=\! N/2 $.

Finally, we summarize the performance of the three considered cases.
Comparing the two cases without temporal mixing, i.e., Cases 1 and 3, we observe that the use of time-variant spatial mixing in Case 3 overcomes the challenge of lack of diversity observed in Case 1 in the extremely sparse case and results in a better estimation quality.
On the other hand, in the case without temporal mixing, the signal in each time-slot is measured independently and, hence, 
from the magnitude-only measurements, the signals can only be recovered up to a global phase ambiguity for each time-slot.
Thus, the temporal mixing, which is applied in Case 2, is introduced to further recover the relative phase between the signals in different time-slots.

\section{Conclusion}
\label{sec:conclusion}
In this paper, we introduce an extension of SCA framework for the phase retrieval with dictionary learning problem. Two efficient parallel algorithms are proposed by applying the extended SCA framework to two complementary formulations, respectively.
The first algorithm, termed \textit{compact-SCAphase}, employs a compact $ \ell_1 $-regularized nonconvex LS formulation, which avoids the auxiliary variables required in state-of-the-art methods such as SC-PRIME and DOLPHIn. The second algorithm, denoted by \textit{SCAphase}, solves the conventional formulation as in SC-PRIME.
An efficient procedure based on rational approximation is devised for solving the $\ell_2$-norm constrained LS subproblems under the SCA framework.
For both algorithms, we refined the search range for suitable values of the sparsity parameter.
Simulation results on synthetic data in the context of blind channel estimation in multi-antenna random access network demonstrate the fast convergence of SCAphase compared to SC-PRIME. Moreover, compact-SCAphase is more competitive than SCAphase in terms of both computational complexity and parameter tuning cost in the case with less diverse linear measurement operators.
Nevertheless, SCAphase also has several advantages over compact-SCAphase. Compared to SCAphase, the computational complexity of compact-SCAphase dramatically grows with the increase of diversity of the designed linear measurement operator. Also, SCAphase can easily include potential side constraints on the signal of interest.

Several questions that have been answered for the classic phase retrieval remain open for phase retrieval with dictionary learning. 
First, further work needs to be done to establish the theoretical conditions for a guaranteed unique recovery (up to trivial ambiguities) of the dictionary and/or the sparse codes.
Moreover, the simulation results in Section~\ref{sec:results} show that multiple random initializations are required for attaining (near\nobreakdash-)global minima of our nonconvex formulations.
Hence, it is of great interest to develop a more sophisticated initialization strategy that can help avoid poor stationary points.

\appendices

\section{Subproblem~\eqref{prob:approx_dp} with $ \opr{F} $ in~\eqref{eq:oprF_time-invariant}}
\label{appendix:subprobD}

For the linear operator $ \opr{F} $ in~\eqref{eq:oprF_time-invariant}, the matrix $ \bF $ in the vectorized form~is
	$ \bF = \bB^\tT \otimes \bA $
and then we have
	$ \bH_p = (\bB^\tT \otimes \bA ) \cdot ( \bz_{p:}^{(t)} \otimes \bI_{N} ) = (\bB^\tT \bz_{p:}^{(t)}) \otimes \bA $.
Let $ \bA = \bU_A \bSigma_A \bV_A^\tH$ and $ \bB^\tT \bz_{p:}^{(t)} = \bU_B \bSigma_B \bV_B^\tH $ be the compact SVDs of $ \bA $ and $ \bB^\tT \bz_{p:}^{(t)} $, respectively. The compact SVD of $ \bH_p $ can be analytically calculated as~\cite{golubMatrixComputations2013}:
\begin{equation*}
	\bH_p = \underbrace{(\bU_B \otimes \bU_A)}_{\bU} \underbrace{(\bSigma_B \otimes \bSigma_A)}_{\bSigma} \underbrace{(\bV_B \otimes \bV_A)^\tH}_{\bV^\tH}.
\end{equation*} 
As a column vector, $ \bB^\tT \bz_{p:}^{(t)} $ has $ \bV_B \!=\! 1 $ and only one nonzero singular value $ \norm{\bB^\tT \bz_{p:}^{(t)}}_2 $. Thus, 
we have 
$ \bU \bSigma = (\bU_B \bSigma_B) \otimes (\bU_A \bSigma_A) = (\bB^\tT \bz_{p:}^{(t)}) \otimes (\bU_A \bSigma_A) $,
and the nonzero singular values of $ \bH_p $ are given by 
	$ \norm{\bB^\tT \bz_{p:}^{(t)}}_2 \cdot \sigma_i^A$, $i \!=\! 1,\ldots,r $,
where $ \{\sigma_i^A\}_{i=1}^r $ are the nonzero singular values of $ \bA $ and $ r \!=\! \rank(\bH_p) \!=\! \rank(\bA) $. 
Consequently, vector $ \bc_p $ in~\eqref{eq:rational} can be written as
\begin{equation*}
	\bc_p \!=\! (\bB^\tT \bz_{p:}^{(t)})^\tH \! \otimes \! (\bU_A \bSigma_A)^\tH \cdot \vec (\bY_p^{(t)}) \!=\! \bSigma_A^\tH \bU_A^\tH  \bY_p^{(t)}  \bB^\tH \bar{\bz}_{p:}^{(t)}.
\end{equation*}
Finally, 
after having obtained the dual optimal solution $ \widetilde{\nu}_p $ by the same procedure as described in Section~\ref{sec:alg1}, we can also compute the optimal solution $ \widetilde{\bd}_p $ using the SVD of $ \bA $:
\begin{equation*}
	\widetilde{\bd}_p= \bV_A \big( \bSigma_A^\tH \bSigma_A + \widetilde{\nu}_p \bI_r \big)^\dagger \bc_p.
\end{equation*}

\section{Proof of Theorem~\ref{thm:rationalLowerBounding}}
\label{appendix:rationalLowerBounding}

The original rational function $\psi (\nu)$ in~\eqref{eq:rational} and its derivative $\psi'(\nu)$ can be rewritten as
\begin{equation*}
	\psi(\nu) = \begin{matrix}
		\sum_{i=1}^r
	\end{matrix} \tfrac{\abs{c_i}^2}{(\delta_i- \nu)^2} \quad \text{and} \quad \psi'(\nu) = \begin{matrix}
	\sum_{i=1}^r
\end{matrix} \tfrac{2\abs{c_i}^2}{(\delta_i - \nu)^3}
\end{equation*}
with the poles $ \delta_1 \leq \cdots \leq \delta_r < 0$. We ignore the trivial case where all poles $\delta_i$ are identical. 
Define $\zeta(\nu) = F (\nu; \alpha,\beta) - \psi(\nu)$ with the approximate function $F$ defined in~\eqref{eq:rationalSimple}. It is sufficient to show that $\zeta(\nu) < 0$ for all $\nu > \delta_r$ and $\nu \neq \nu^{(l)}$. To this end, define
\begin{equation*}
	\begin{matrix}
		\xi(\nu) = \zeta(\nu) (\beta - \nu)^2 \prod_{i=1}^{r} (\delta_i - \nu)^2
	\end{matrix}.
\end{equation*}
Then $\xi$ is a polynomial of degree $2r$ with real coefficients:
\begin{equation}\label{eq:xi_poly}
		\xi(\nu) = \alpha \prod_{i=1}^{r} (\delta_i - \nu)^2 - (\beta - \nu)^2 \sum_{i=1}^r \abs{c_i}^2 \prod_{j=1, j\neq i}^r (\delta_j - \nu)^2.
\end{equation}
The product rule for differentiation determines that $\xi(\nu^{(l)}) = 0$ and its derivative $\xi'(\nu^{(l)}) = 0$ since $\zeta(\nu^{(l)}) = 0$ and its derivative $\zeta'(\nu^{(l)}) = 0$. Hence, $\nu^{(l)}$ is a double root of $\xi$, and we can extract the factor $(\nu - \nu^{(l)})^2$ and rewrite~\eqref{eq:xi_poly} as
\begin{equation*}
	\xi(\nu) = \big(\alpha - \begin{matrix}
		\sum_{i=1}^r
	\end{matrix} \abs{c_i}^2\big) \big(\nu - \nu^{(l)} \big)^2 \begin{matrix}
	\prod_{i=1}^{r-1}
\end{matrix} (\nu^2 - 2a_i \nu + b_i)
\end{equation*}
with appropriately chosen coefficients $a_i, b_i \in \Rbb$.

\begin{figure}[t]
	\centering
	\input{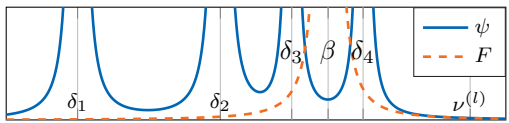}
	\caption{Original and approximate rational functions, $r=4$.}
	\label{fig:rationalLowerBounding}
\end{figure}

We claim that $\nu^{(l)}$ is the only real double root of $\xi$ in $(\delta_r, + \infty)$.
To see this, observe from~\eqref{eq:alphabeta} that the pole of $ F $
	$ \beta = \tfrac{2}{\psi'(\nu^{(l)})} \begin{matrix}
		\sum_{i=1}^r
	\end{matrix}  \tfrac{ \delta_i\abs{c_i}^2}{(\delta_i - \nu^{(l)})^3} \in (\delta_1, \delta_r) $.
The roots of $\zeta$ are also the roots of $\xi$. 
The following result can be intuitively observed from Fig.~\ref{fig:rationalLowerBounding}. Each interval $ (\delta_i, \delta_{i+1}) $ with $ \delta_i \neq \delta_{i+1} $ contains either two real roots of $\zeta$ or the real part of a pair of complex conjugate roots.
In contrast, if $ \delta_i = \delta_{i+1} $ for some $i = 1,\ldots,r-1$, it can be trivially identified from~\eqref{eq:xi_poly} that $\delta_i$ is a double root of $\xi$. Hence, the real parts of the remaining $2r-2$ roots of $\xi$ fall in the interval $ [\delta_1, \delta_r] $. The claim is established; it can be proved more formally by factorizing~\eqref{eq:xi_poly}.

This argument shows that $\sign (\xi (\nu))$ remains constant in $[\delta_r,\nu^{(l)}) \cup (\nu^{(l)}, + \infty)$. Therefore, it follows from~\eqref{eq:xi_poly} that, for all $\nu > \delta_r$ and $\nu \neq \nu^{(l)}$,
	$ \sign (\zeta (\nu)) = \sign (\xi (\nu)) = \sign (\xi(\delta_r))  
	= -1 $.
This implies that $F(\nu;\alpha,\beta) < \psi(\nu)$ for all $\nu > \delta_r$ and $\nu \neq \nu^{(l)}.$ \hfill $ \blacksquare $

\section{Proof of Theorem~\ref{prop:convergence}}
\label{appendix:convergence}

In this paper, Algorithm~\ref{alg1} and Algorithm~\ref{alg2} are developed for the phase retrieval with dictionary learning problem. However, this framework can be easily generalized to another nonconvex nonsmooth problem with a continuous and locally Lipschitz objective function, and the convergence is ensured under several assumptions on the approximate functions. Therefore, we first demonstrate the convergence of the generalized algorithm for the general constrained problem~\eqref{prob:generalConstrained}. Then we verify that the required assumptions are satisfied in the compact-SCAphase algorithm and, thus, Theorem~\ref{prop:convergence} can be proved.
In the following analysis, for simplicity, we ignore the convex nonsmooth regularization, e.g., the $\ell_1$-regularization in~\eqref{prob:PRDLMatrixForm}. 
However, a regularized problem can be written in the standard form~\eqref{prob:generalConstrained} with the reformulation in~\cite[Eq. (18)]{yangUnifiedSuccessivePseudoconvex2017} and then the analysis below can be directly applied. 

Let $ \widehat{f} (\bs;\vect{w}) $ be a smooth majorizing function of $f$ in~\eqref{prob:generalConstrained} at $\vect{w} \in \set{C} $ and $ \widetilde{f} (\bs;\vect{w}) $ be a pseudoconvex approximation of $ \widehat{f} (\vect{s}; \vect{w}) $ at the same point $ \vect{w} $ in the generalized algorithm.
Precisely, $ \widehat{f} $ and $ \widetilde{f} $ are constructed to satisfy the following assumptions:

\textbf{(A1)} $ \widetilde{f} (\bs; \vect{w}) $ is pseudoconvex in $ \bs \in \set{C} $ for any $ \bw \in \set{C} $;

\textbf{(A2)} $ \widetilde{f} (\bs; \vect{w}) $ and $ \widehat{f} (\bs; \vect{w}) $ are $ C^1 $-smooth in $\vect{s} \in \set{C} $ for any $ \vect{w} \in \set{C} $ and continuous in $\vect{w} \in \set{C}$ for any $ \vect{s} \in \set{C} $;

\textbf{(A3)} $ \widehat{f} (\bs; \vect{w}) \geq f(\bs) \ \forall \bs, \vect{w} \in \set{C}$ and $ \widehat{f} (\vect{w};\vect{w}) = f(\vect{w}) \ \forall \vect{w} \in \set{C}$;

\textbf{(A4)} $ \nabla_{\vect{s}} \widetilde{f} (\vect{w};\vect{w}) = \nabla_\vect{s} \widehat{f} (\vect{w};\vect{w}) \in \partial_C f(\vect{w}) \ \forall \vect{w} \in \set{C} $;

\textbf{(A5)} $ \widetilde{f} (\vect{s}; \vect{s}^{(t)}) $ has an attainable minimizer in $ \set{C} $ for $ t \in \mathbb{N}$, and the sequence $ (\widetilde{\vect{s}}^{(t)})_t $ is bounded.

Then, under assumptions (A1)-(A5), the solution sequence obtained by the generalized algorithm converges to a C-stationary point of problem~\eqref{prob:generalConstrained}.
The proof is as follows.
The convergence analyses of the MM algorithms~\cite{sunMajorizationMinimizationAlgorithmsSignal2017} and the SCA framework~\cite{yangUnifiedSuccessivePseudoconvex2017} for a smooth function $f$ are actually equivalent in the sense that they both rely on the two essential facts
corresponding to the two cases where the current point $\bs^{(t)}$ is a fixed point and where $\bs^{(t)}$ is not a fixed point, respectively:
\begin{enumerate}[\textit{Fact} 1:]
	\item \label{fact1:stationary} $\bs^{(t)}$ is a fixed point, i.e., a stationary point of the majorizing/approximate problem, if and only if it is a stationary point of the original problem;
	\item \label{fact2:descent} otherwise, if $ \bs^{(t)} $ is not a fixed point, then a strict decrease of $f$ is obtained through the solution of the majorizing/approximate problem.
\end{enumerate}
Similarly, in the following, we first justify that Facts~\ref{fact1:stationary} and~\ref{fact2:descent}, with the generalized concept of stationarity, hold for our proposed algorithm so as to demonstrate the convergence.
First, if $\bs^{(t)}$ is not a fixed-point, assumptions (A1), (A2), and (A4) ensure that the minimizer of $ \widetilde{f} $ indicates a descent direction of the majorizing function $ \widehat{f} $. Then a decrease of the original function $f$ is achieved in our proposed algorithm by exact line search on $ \widehat{f} $ in this descent direction:
\begin{equation}\label{eq:descentProperty}
	f(\bs^{(t+1)}) \leq \widehat{f} (\bs^{(t+1)}; \vect{s}^{(t)}) < \widehat{f} (\bs^{(t)}; \vect{s}^{(t)}) = f(\bs^{(t)}).
\end{equation}
Second, to avoid the exact minimization of the majorizing function $ \widehat{f} (\vect{s};\vect{s}^{(t)}) $, which is required in the classic MM algorithm, our algorithm develops a pseudoconvex approximation $ \widetilde{f}(\vect{s};\vect{s}^{(t)}) $ that is easier to minimize and retains the gradient of $ \widehat{f}(\vect{s};\vect{s}^{(t)}) $ at $\bs^{(t)}$.
Specifically, from the smoothness assumption (A2) and subgradient consistency assumption (A4), we have
\begin{align} 
	\label{eq:subgradientConsistency}
	\partial_C \widetilde{f} (\vect{w};\vect{w}) + \set{N}_{\set{C}} (\vect{w}) &= \partial_C \widehat{f} (\vect{w}; \vect{w}) + \set{N}_{\set{C}} (\vect{w}) \nonumber \\
	&\subseteq \partial_C f(\vect{w}) + \set{N}_{\set{C}} (\vect{w})
\end{align}
with
$ \partial_C \widetilde{f}(\vect{w};\vect{w}) \!=\! \{\nabla_{\vect{s}} \widetilde{f} (\vect{w};\vect{w})\} $ and
$ \partial_C \widehat{f} (\vect{w};\vect{w}) \!=\! \{ \nabla_{\vect{s}} \widehat{f}(\vect{w};\vect{w})\} $ for any $ \vect{w} \in \set{C} $ \cite{clarkeOptimizationNonsmoothAnalysis1990}.
By the definition of C-stationarity in~\eqref{eq:C-stationarity}, \eqref{eq:subgradientConsistency} justifies that Fact~\ref{fact1:stationary} holds for our proposed algorithm. In other words, the minimization of $ \widetilde{f} (\vect{s}; \vect{s}^{(t)}) $ is sufficient for determining whether $ \bs^{(t)} $ is a C-stationary point of $ f $ and, hence, the minimization of $\widehat{f} (\vect{s};\vect{s}^{(t)})$ is not required.

If a fixed point is achieved in a finite number of iterations, then Fact~\ref{fact1:stationary} ensures the convergence to a C-stationary point of the original problem. Otherwise, by following the same procedures as in~\cite{yangUnifiedSuccessivePseudoconvex2017}, we show that $ (\vect{s}^{(t)})_t $ asymptotically converges to a C-stationary point of the original problem for $ t \rightarrow \infty $, based on Fact~\ref{fact2:descent}. Fact~\ref{fact2:descent} implies that $\big( f (\vect{s}^{(t)}) \big)_t $ is a monotonically decreasing sequence, which, by the monotone convergence theorem, converges to a local minimum of $f$ in $ \set{C} $, assuming that $ f $ is bounded below in $ \set{C} $.
Thus, for any two convergent subsequences $(\vect{s}^{(t)})_{t \in \set{T}_1 \subseteq \mathbb{N}}$ and $ (\vect{s}^{(t)} )_{t \in \set{T}_2 \subseteq \mathbb{N}} $, it holds that
	$ \lim_{t \rightarrow \infty} f(\vect{s}^{(t)}) = \lim_{t \in \set{T}_1, t \rightarrow \infty} f(\vect{s}^{(t)}) = \lim_{t \in \set{T}_2, t \rightarrow \infty} f(\vect{s}^{(t)}) $.
Since $f(\vect{s})$ is a continuous function, it follows that
\begin{equation}\label{eq:limit}
	\begin{matrix}
		f \big( \lim_{t \in \set{T}_1, t \rightarrow \infty} \vect{s}^{(t)} \big) = f \big( \lim_{t \in \set{T}_2,  t \rightarrow \infty} \vect{s}^{(t)} \big).
	\end{matrix}
\end{equation}
Now consider a convergent sequence $(\vect{s}^{(t)})_t$ with limit point $\vect{z} \in \set{C}$, i.e., $ \lim_{t \rightarrow \infty} \vect{s}^{(t)} = \vect{z} $. Let $\widetilde{\vect{z}}$ be one minimizer of the approximate function $\widetilde{f}(\vect{s};\vect{z})$ and define the set $ \set{S} (\vect{z}) = \{ \widetilde{\vect{z}} \mid \widetilde{\vect{z}} = {\arg \min}_{\vect{s} \in \set{C}} \ \widetilde{f}(\vect{s};\vect{z}) \}. $
Under the assumptions that $\widetilde{f}(\vect{s};\vect{w})$ is continuous in both $\vect{s}$ and $ \vect{w} $, and that $(\widetilde{\vect{s}}^{(t)})_t$ is bounded, it follows from the maximum theorem~\cite[Sec. VI.3]{bergeTopologicalSpacesIncluding1997} that there exists a convergent subsequence $ (\widetilde{\vect{s}}^{(t)})_{t \in \set{T}_s \subseteq \mathbb{N}} $ with $ \lim_{t \in \set{T}_s, t \rightarrow \infty} \widetilde{\vect{s}}^{(t)} \in \set{S}(\vect{z}) $. Further applying the maximum theorem on the exact line search problem implies that there exists a subsequence $ (\vect{s}^{(t+1)})_{t \in \set{T}_{s'} \subseteq \set{T}_s} $ that converges to $\vect{z}'$ defined as $ \vect{z}' = \vect{z} + \gamma (\widetilde{\vect{z}} - \vect{z}) $, where $\gamma$ is the step size obtained by the exact line search on $ \widehat{f} (\vect{s}; \vect{z}) $ at $\vect{z}$ in the direction $ \widetilde{\vect{z}} - \vect{z} $. If $\vect{z}$ is not a C-stationary point of $f$, which, by~\eqref{eq:subgradientConsistency}, is neither a stationary point of the approximate function $ \widetilde{f} (\vect{s};\vect{z}) $, then Fact 2 implies that $f(\vect{z}') < f(\vect{z})$, which contradicts~\eqref{eq:limit}. Therefore, any limit point of $ (\vect{s}^{(t)})_t $ is a C-stationary point of the original problem. 



Moreover, provided that assumptions (A2) and (A3) are satisfied, assumption (A4) is satisfied under the following assumption that is easier to verify:

\textbf{(A6)} $ \nabla \widetilde{f} (\vect{w};\vect{w}) = \nabla \widehat{f} (\vect{w};\vect{w}) $ for any $\vect{w} \in \set{C}$ and $f(\bs)$ is directionally differentiable for all $\bs \in \set{C}$.

In other words, the convergence of the generalized algorithm is also ensured under assumptions (A1)-(A3), (A5), and (A6).
The proof is as follows. Under the smoothness assumption (A2), the directional derivative of $ \widehat{f} (\vect{s};\vect{w}) $ in any direction $ \vect{r} \in \Rbb^n $ is given by
	$ \widehat{f}' (\bs; \vect{w},\br) = \br^\tT \nabla_{\vect{s}} \widehat{f} (\bs,\vect{w})\ \forall \bs,\vect{w} \in \set{C}$.
As $f$ is directionally differentiable, the majorization assumption (A3) implies that
	$ \br^\tT \nabla_{\vect{s}} \widehat{f} (\vect{w};\vect{w}) \geq f'(\vect{w}; \br) \ \forall \br \in \Rbb^n $.
It follows that, for all $ \vect{w} \in \set{C} $ and $ \vect{r} \in \Rbb^n$,
\begin{equation}\label{eq:direcDerivBoud}
	- \br^\tT \nabla_{\vect{s}} \widehat{f} (\vect{w};\vect{w}) \leq - f' (\vect{w}; \br) \leq f^\circ (\vect{w}; -\br), 
\end{equation}
where the last inequality comes from the definition of Clarke directional derivative in~\eqref{eq:C-direcDeriv}. By the definition of C-subdifferential in~\eqref{eq:C-subdiff}, we conclude from~\eqref{eq:direcDerivBoud} that $ \nabla_{\vect{s}} \widehat{f} (\vect{w};\vect{w}) \in \partial_C f(\vect{w}) $ and, consequently, assumption (A4) is satisfied.

Now, for the compact-SCAphase algorithm that solves~\eqref{prob:PRDLMatrixForm}, it is trivial to verify that the assumptions (A1)-(A3), (A5), and (A6) are satisfied. Consequently, the solution sequence generated by compact-SCAphase converges to a C-stationary point of problem~\eqref{prob:PRDLMatrixForm} and Theorem~\ref{prop:convergence} is proved.\hfill $ \blacksquare $

\section{Derivation of Upper Bound $\lambda_{\max}$}
\label{appendix:lambda_max}

We derive the upper bound $\lambda_{\max}$ for the sparsity parameter $\lambda$ in~\eqref{prob:PRDLMatrixForm} using the stationarity conditions~\eqref{eq:stationaryCondition} with the gradients in~\eqref{eq:grad_upperBoundf}. Condition~\eqref{eq:stationaryConditionD} is trivial for ${\bZ} \!=\! \bzero$ as $\nabla_{\bD} \widehat{f} ({\bD}, \bzero;\vect{S}^{(t)}) \!=\! \bzero$ for any ${\bD}$. 
Then, adopting the vectorized form in~\eqref{eq:linearOpr_vectorized} for $\opr{F}$ and the partition in~\eqref{eq:partition_F}, we can write the gradient $ \nabla_{\! z_{p,i}} \widehat{f} $ at $\bZ = \bzero$ as
$ 	\nabla_{\! z_{p,i}} \widehat{f} ({\bD}, \bzero;\vect{S}^{(t)}) = - {\bd}_p^{\tH} \bF_i^\tH \vec (\bY^{(t)}) $.
It follows that
\begin{subequations}
	\label{eq:grad_zpi_upperBound}
	\begin{align}
		\abs{\nabla_{\! z_{p,i}} \widehat{f} ({\bD}, \bzero; \vect{S}^{(t)}) } 
		& {\leq} \norm{\bF_i^\tH \vec (\bY^{(t)})}_2 \label{eq:grad_zpi_b}\\
		& {\leq} \norm{\bF_i}_2  \norm{\bY}_\tF, \label{eq:grad_zpi_c}
	\end{align}
\end{subequations}
where~\eqref{eq:grad_zpi_b} comes from the Cauchy--Schwartz inequality and the constraint of problem~\eqref{prob:PRDLMatrixForm},
and the matrix $\ell_2$-norm $\norm{\bF_i}_2$ is equal to the largest singular value of $\bF_i$, denoted by $ \sigma_{\max} (\bF_i) $. Inequality~\eqref{eq:grad_zpi_upperBound} holds for any solution with $\bZ = \bzero$. Consequently, comparing~\eqref{eq:grad_zpi_upperBound} with~\eqref{eq:stationaryConditionZ} yields the following result. Define
\begin{equation*}
	\begin{matrix}
		\lambda_{\max} = \norm{\bY}_\tF \cdot \max_{i=1,\ldots,I} \{ \sigma_{\max}(\bF_i) \} 
	\end{matrix}.
\end{equation*}
For $\lambda \! \geq \! \lambda_{\max}$, any point $(\bD,\bzero)$ with $\bD \in \set{D}$ satisfies the conditions~\eqref{eq:stationaryCondition} and, therefore, is stationary for $\widehat{h}$ in the domain of problem~\eqref{prob:PRDLMatrixForm}. Note that $\lambda_{\max}$ above does not depend on the point $\vect{S}^{(t)}$ where the majorization is made. Hence, $(\bD,\bzero)$ is stationary for $\widehat{h}$ taken at any point, including $(\bD,\bzero)$. 
This implies that, for $ \lambda \geq \lambda_{\max} $, any point $(\bD,\bzero)$ is stationary for the original problem~\eqref{prob:PRDLMatrixForm}.
Also, it is easy to verify that all points $(\bD,\bzero)$ with $\bD \! \in \! \set{D}$ are equally optimal for both $\widehat{h}$ and $h$. 

In addition, $\lambda_{\max}$ can be further decreased in the investigated cases 1 and 2 in Section~\ref{sec:results}, 
where the linear operator $ \opr{F} $ is given by \eqref{eq:oprF_time-invariant}.
In this case,
we have, for $ i=1,\ldots,I, $
\begin{equation}\label{eq:F_i}
	\bF_i = \bb_{i:} \otimes \bA \quad \text{and} \quad \bF_i^\tH \vec (\bY^{(t)}) = \bA^\tH \bY^{(t)} \bar{\bb}_{i:}.
\end{equation}
Then, directly substituting $\bF_i$ in~\eqref{eq:F_i} into~\eqref{eq:grad_zpi_c}, we obtain
\begin{equation} \label{eq:grad_zpi_upperBound_specialF}
	\abs{\nabla_{\! z_{p,i}} \widehat{f} ({\bD}, \bzero; \vect{S}^{(t)}) } \leq \norm{\bA}_2 \norm{\bb_{i:}}_2 \norm{\bY}_\tF.
\end{equation}
On the other hand, exploiting the structure of $\bF_i$ in~\eqref{eq:F_i}, we can further derive the following inequality from~\eqref{eq:grad_zpi_b}:
\begin{multline} \label{eq:grad_zpi_upperBound_refined}
	\abs{\nabla_{\! z_{p,i}} \widehat{f} ({\bD}, \bzero;\vect{S}^{(t)}) }
	\leq \norm{\bA^\tH \bY^{(t)} \bar{\bb}_{i:}}_2 {\leq} \norm{\bA}_2  \norm{\bY^{(t)} \bar{\bb}_{i:}}_2 \\ 
	 = \begin{matrix}
			\norm{\bA}_2 \cdot \big\|{\sum_{m=1}^{M_2} \bar{b}_{i,m} \by_m^{(t)}} \big\|_2 
		\end{matrix}
	{\leq} \norm{\bA}_2 \cdot \begin{matrix}
		 \sum_{m=1}^{M_2}
	\end{matrix} \Abs{b_{i,m}} \cdot \norm{\by_m}_2.
\end{multline}
It is shown by Cauchy--Schwartz inequality that \eqref{eq:grad_zpi_upperBound_refined} is a tighter bound for $\nabla_{\! z_{p,i}} \widehat{f}$ than \eqref{eq:grad_zpi_upperBound_specialF}.
Consequently, in the case with $\opr{F}(\bX) = \bA \bX \bB$, the upper bound $\lambda_{\max}$ can be decreased to
\begin{equation*}
	\begin{matrix}
		\lambda_{\max} = \sigma_{\max}(\bA) \cdot \max_{i=1,\ldots,I} \big\{ \sum_{m=1}^{M_2} \Abs{b_{i,m}} \cdot \Norm{\by_m}_2 \big\}
	\end{matrix}.
\end{equation*}

Furthermore, in Case 3 in Section~\ref{sec:results}, where spatial mixing is time-variant and temporal mixing is not applied, we have
	$ \bF_i = \be_i \otimes \bA_i $
from the vectorized form in~\eqref{eq:oprF_withoutTemporalMixing}.
Therefore, following the same procedure as in~\eqref{eq:grad_zpi_upperBound_refined}, we obtain the following bound for $\nabla_{\! z_{p,i}} \widehat{f}$ tighter than~\eqref{eq:grad_zpi_c}:
	$ \abs{\nabla_{\! z_{p,i}} \widehat{f} ({\bD}, \bzero) } \leq \norm{\bA_i}_2 \norm{\by_i}_2 $.
Consequently, in Case 3, $\lambda_{\max}$ can be refined to
\begin{equation*}
	\begin{matrix}
		\lambda_{\max} = \max_{i=1,\ldots,I} \left\{ \sigma_{\max} (\bA_i) \cdot \norm{\by_i}_2 \right\}
	\end{matrix}.
\end{equation*}

\section{Derivation of Upper Bound $\rho_{\max}$}
\label{appendix:rho_max}

We derive the upper bound $\rho_{\max}$ for the sparsity parameter $\rho$ in~\eqref{prob:SC-PRIME} using the stationarity conditions~\eqref{eq:stationaryCondition_2}-\eqref{eq:stationaryConditionX1} with the gradients in~\eqref{eq:grad_upperBoundf_2}. 
Condition~\eqref{eq:stationaryConditionD_2} is trivial for ${\bZ} \!=\! \bzero$ as, for any $\bX$ and ${\bD}$, $\nabla_{\bD} \widehat{f} (\bX, {\bD}, \bzero;\vect{S}^{(t)}) \!=\! \bzero$. 
As for~\eqref{eq:stationaryConditionZ_2}, we have
\begin{equation}
	\hspace*{-4pt} \abs{\nabla_{\!\! z_{p,i}} \widehat{f} (\bX,\bD,\bzero;\vect{S}^{(t)})} \!=\! \mu \abs{\bd_p^\tH \bx_i} \! \leq \! \mu \norm{\bd_p}_2 \norm{\bx_i}_2 \! \leq \! \mu \norm{\bx_i}_2.  \label{eq:upperBound_gradZ_2}
\end{equation}
Meanwhile, an upper bound for $\norm{\bx_i}_2$ can be derived from the vectorized form~\eqref{eq:stationaryConditionX1_2} of condition~\eqref{eq:stationaryConditionX1},
which reduces to
\begin{equation}
	\label{eq:stationaryConditionX1_Z=0}
	\vec (\bX) = ( \bF^\tH \bF + \mu \bI_{NI} )^{-1} \bF^\tH \vec (\bY^{(t)}),
\end{equation}
for $ \bZ = \bzero $.
It leads to the following upper bound for $ \norm{\bx_i}_2 $:
\begin{equation} \label{eq:upperBound_X}
	\norm{\bx_i}_2 
	\leq \norm{\bX}_\tF 
	\leq \norm{( \bF^\tH \bF + \mu \bI_{NI} )^{-1}}_2 \norm{\bF}_2 \norm{\bY}_\tF.
\end{equation}
As an oversampling operator $\opr{F}$ is considered, i.e., $M_1 M_2 \geq NI$, we have
	$ \norm{( \bF^\tH \bF \!+\! \mu \bI_{NI} )^{-1}}_2 = \big({\sigma_{\min}^2(\bF) \!+\! \mu }\big)^{-1} $.
Consequently, combining~\eqref{eq:upperBound_gradZ_2} and~\eqref{eq:upperBound_X} yields the following result. Define
\begin{equation}
	\label{eq:rho_max_general_2}
	\begin{matrix}
		\rho_{\max} = {\mu \cdot \sigma_{\max}(\bF) \cdot \norm{\bY}_\tF} \big/ {\big(\sigma_{\min}^2(\bF)+ \mu\big)}
	\end{matrix}.
\end{equation}
For $\rho \geq \rho_{\max}$, there always exists a feasible point $(\bX,\bD,\bzero)$ that satisfies the stationarity conditions~\eqref{eq:stationaryCondition_2} and~\eqref{eq:stationaryConditionX1}, and is, therefore, stationary for the majorizing function $\widehat{h}$. As $\rho_{\max}$ does not depend on the point $\vect{S}^{(t)}$ where the majorization is made, following the same line of arguments as in Appendix~\ref{appendix:lambda_max}, we further conclude that, for any $\rho \geq \rho_{\max}$, the original problem~\eqref{prob:SC-PRIME} admits a stationary point with $\bZ = \bzero$.

In the investigated cases 1 and 3 in Section~\ref{sec:results}, where temporal mixing is not applied, the linear operator $\opr{F}$ and the corresponding matrix $\bF$ can be expressed as
\begin{equation}
	\label{eq:oprF_withoutTemporalMixing}
	\begin{matrix}
		\opr{F}(\bX) = \sum_{i=1}^I \bA_i \bX \bB_i \quad \text{and} \quad \bF = \sum_{i=1}^I \bB_i^\tT \otimes \bA_i
	\end{matrix},
\end{equation}
where $\bB_i \!=\! [\bzero,\ldots,\bzero,\be_i,\bzero,\ldots,\bzero]$ selects the $i$-th snapshot $\bx_i$ and $\bA_i$ is the spatial mixing designed for $\bx_i$.
In Case 1, where the spatial mixing is time-invariant, all matrices $\bA_i$ are set to be the same value $\bA$.
Substituting $\bF$ in~\eqref{eq:oprF_withoutTemporalMixing} into~\eqref{eq:rho_max_general_2}, we obtain
\begin{equation*}
	\rho_{\max} = \tfrac{\mu \cdot \max_{i=1,\ldots,I}\{ \sigma_{\max}(\bA_i)\}}{\min_{i=1,\ldots,I} \{ \sigma_{\min}^2(\bA_i)\} + \mu } \cdot \norm{\bY}_\tF.
\end{equation*}
However, the upper bound $\rho_{\max}$ can be further decreased considering that each snapshot $\bx_i$ is observed independently when temporal mixing is not applied.
Using the matrix $\bF$ in~\eqref{eq:oprF_withoutTemporalMixing}, we can reformulate the stationary condition~\eqref{eq:stationaryConditionX1_Z=0} as
\begin{equation*}
	\bx_i = (\bA_i^\tH \bA_i + \mu \bI_N)^{-1} \bA_i^\tH \by_i^{(t)} \quad \text{for } i=1,\ldots,I.
\end{equation*}
This results in a tighter bound of $ \norm{\bx_i}_2 $ than~\eqref{eq:upperBound_X}:
\begin{equation*}
	\norm{\bx_i}_2 \leq \norm{(\bA_i^\tH \bA_i + \mu \bI_N)^{-1}}_2 \norm{\bA_i}_2 \norm{\by_i^{(t)}}_2.
\end{equation*}
Thus, in Cases 1 and 3, the upper bound $\rho_{\max}$ is refined to
\begin{equation*}
		\rho_{\max} = \max_{i=1,\ldots,I} \big\{{\mu \cdot \sigma_{\max}(\bA_i)\cdot \norm{\by_i}_2} \big/ {\big(\sigma^2_{\min} (\bA_i) + \mu\big)}  \big\}.
\end{equation*}

\section*{Acknowledgments}
The authors thank three reviewers for detailed comments that helped to improve the presentation of the paper.
Extensive calculations on the Lichtenberg high-performance computer of TU Darmstadt were conducted for this research. 

\bibliographystyle{IEEEtran}
\bibliography{refsPhaseRetrieval}

\end{document}